\documentclass[aps,prb,twocolumn,superscriptaddress,longbibliography]{revtex4-2}

\usepackage[utf8]{inputenc}
\usepackage[T1]{fontenc}
\usepackage{amsmath,amssymb}
\usepackage{graphicx}
\usepackage{xcolor}
\usepackage[colorlinks=true,linkcolor=blue,citecolor=blue,urlcolor=blue]{hyperref}

\begin{document}

\title{Sector-resolved non-Bloch topology and nonlocal entanglement dynamics in a bond-dissipative Kitaev chain}

\author{Shaina Gandhi}
\email{shaina007@iitg.ac.in}
\affiliation{Department of Physics, Indian Institute of Technology Guwahati, Guwahati 781039, Assam, India}

\author{Koustav Roy}
\email{koustav.roy@iitg.ac.in}
\affiliation{Department of Physics, Indian Institute of Technology Guwahati, Guwahati 781039, Assam, India}

\author{Bilal Tanatar}
\email{tanatar@fen.bilkent.edu.tr}
\affiliation{Department of Physics, Bilkent University, 06800 Bilkent, Ankara, Türkiye}

\author{Saurabh Basu}
\email{saurabh@iitg.ac.in}
\affiliation{Department of Physics, Indian Institute of Technology Guwahati, Guwahati 781039, Assam, India}

\begin{abstract}
A core characteristic of dissipative non-Hermitian topology is that the relaxation dynamics
tracks the non-Bloch bulk-boundary correspondence, rendering an algebraic decay in the
gapless regime and an exponential falloff in the gapped phase, so that local observables
directly diagnose the topology. We show that this correspondence breaks down in a
dissipative topological superconductor, where the local observables turn blind to the very
topology they are expected to decipher. Via a bond-dissipative dimerized Kitaev chain in a
third-quantized rapidity-matrix formulation, we find that at zero chemical potential the
Majorana rapidity matrix decomposes into two independent non-Hermitian sectors, each with
its own generalized Brillouin zone and non-Bloch winding number, thereby revealing a
sector-resolved non-Bloch bulk-boundary correspondence. The local density is a cross-sector
covariance and relaxes at the sum of the two sector rates, so it remains sector-blind even
when one sector is gapless and topological. For balanced gain and loss, the finite-time zero
events of the entanglement spectrum under purely periodic-boundary Lindblad evolution
recover this hidden edge content sector by sector, serving as a dynamical invariant that
returns the open-boundary edge rapidities without physically opening the chain.
\end{abstract}

\maketitle

\emph{Introduction.---}
Non-Hermitian lattices can be strongly boundary sensitive, with open boundaries reshaping the spectrum and causing the bulk eigenstates to accumulate at the edges, a phenomenon known as the non-Hermitian skin effect (NHSE). The resulting failure of Bloch bulk-boundary correspondence motivates a non-Bloch formulation based on generalized Brillouin zones (GBZs) \cite{PhysRevLett.121.086803,PhysRevLett.121.026808,PhysRevLett.123.066404,PhysRevLett.124.086801,PhysRevB.111.115424,s3b6-wz16}. Non-Hermitian topological phases are accordingly classified by the point-gap and line-gap structure of the complex spectrum, the former capturing spectral winding and closely associated with the NHSE, the latter permitting Hermitian-like classifications \cite{PhysRevLett.120.146402,PhysRevX.8.031079,PhysRevX.9.041015,RevModPhys.93.015005,Ashida2020,OkumaSato2023}.

Open quantum systems provide a natural setting for broader manifestations of this boundary-sensitive physics. For Markovian dynamics generated by a quadratic Lindblad master equation \cite{Gorini1976,Lindblad1976,roy2025}, two-point relaxation is governed by a finite-dimensional rapidity (damping) matrix obtained from third quantization \cite{Prosen2008,Prosen2010}. The slowest rapidity decay rate defines the Liouvillian damping gap, and the rapidity matrix itself can display a skin effect, so that its open boundary condition (OBC) spectrum is not obtained from the Bloch unit circle. The resulting gap-relaxation correspondence is well established in the dissipative Su-Schrieffer-Heeger (SSH) chain. There the periodic boundary condition (PBC) damping gap vanishes for $t_1\le t_2$, and the particle density inherits the decay of the gapless channel and relaxes algebraically, whereas under OBC the non-Bloch spectrum is gapped and the skin effect produces a directional damping front, known as chiral damping, beyond which the decay becomes exponential \cite{PhysRevLett.123.170401,PhysRevLett.127.070402,PhysRevResearch.4.023160,PhysRevB.105.064302, PhysRevB.108.054313}. 

We show that this scenario demands an overhaul when the system under consideration is a topological superconductor, hosting Majorana zero modes protected by particle-hole symmetry.  Superconducting pairing couples normal and anomalous correlations, so the closed dynamical object is the full Majorana covariance matrix. Distinct Majorana rapidity channels can then coexist, while a physical observable may mix them rather than resolve them independently. Consequently, a gapless sector need not govern every observable. In the present model, because the local density is built from cross-sector covariances, its relaxation can remain exponential even when one sector is gapless, an insensitivity we call \textit{sector-blindness}. Although dissipative superconductors can host decaying Majorana modes, dissipation-generated topology, and dynamical edge-mode signatures \cite{Diehl2011,PhysRevLett.109.130402,Bardyn_2013,SciPostPhys.6.2.026,SciPostPhys.17.2.036}, it remains unclear how inequivalent Majorana sectors reshape the relation among Liouvillian skin physics, non-Bloch topology, damping gaps, and observable relaxation.

\begin{figure}[t]
    \centering
    \includegraphics[width=0.47\textwidth]{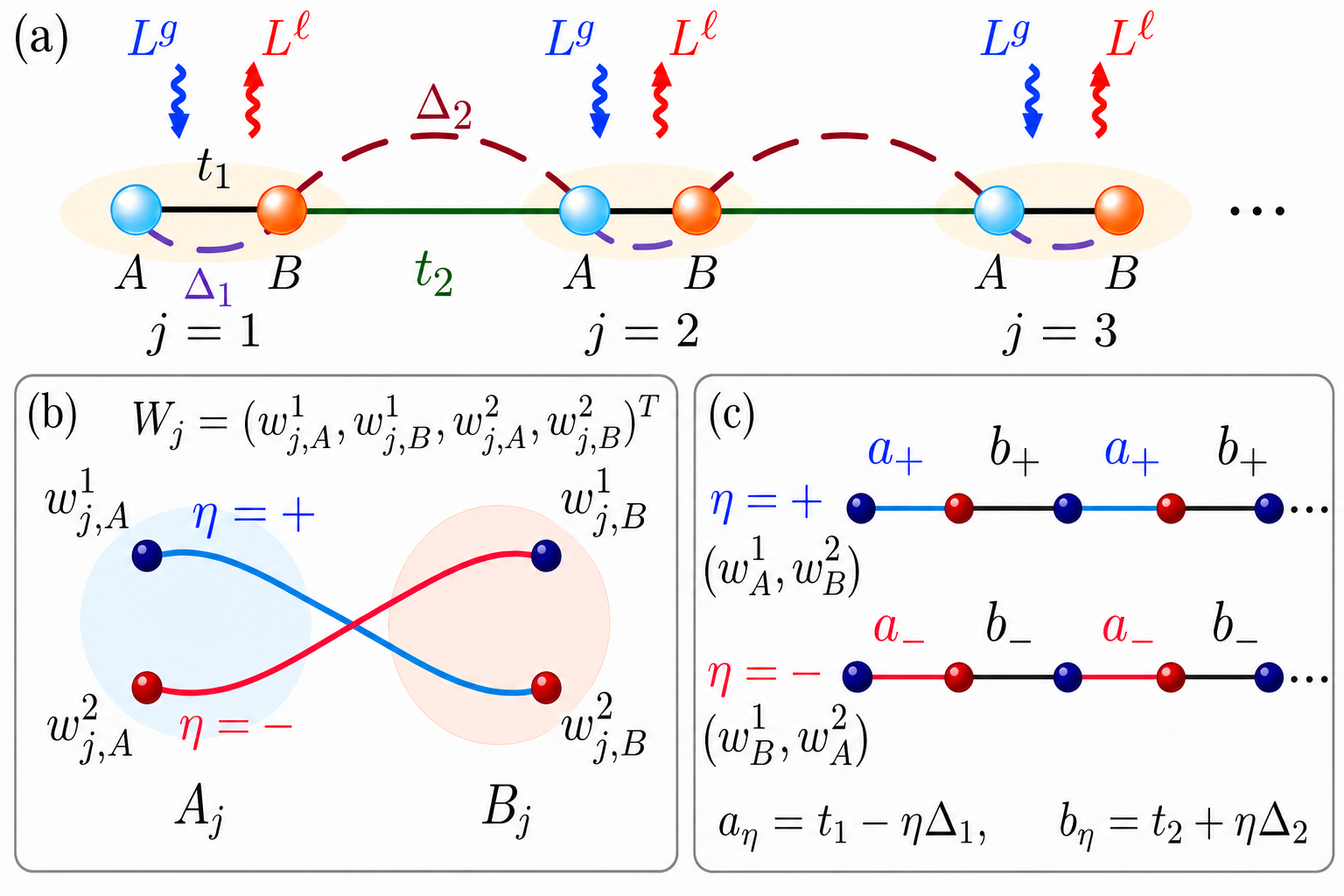}
    \caption{Schematic of the bond-dissipative dimerized Kitaev chain and its Majorana-sector decomposition. (a) Real-space chain with sublattices \(A\) and \(B\) in each unit cell. Solid bonds denote hoppings \(t_1=t(1+d)\), \(t_2=t(1-d)\), dashed bonds represent the pairings \(\Delta_1=\Delta(1-d)\), \(\Delta_2=\Delta(1+d)\), and each unit cell couples to a Markovian bath through the bond loss and gain jump operators \(L^\ell_j\) and \(L^g_j\). (b) Local Majorana basis and sector split at \(\mu=0\), with sector $1\!-\!4$ spanned by \((w^1_A,w^2_B)\) and sector $2\!-\!3$ by \((w^1_B,w^2_A)\). (c) The sectors map to SSH-like Majorana chains with effective couplings \(a_\eta=t_1-\eta\Delta_1\) and \(b_\eta=t_2+\eta\Delta_2\).}
    \label{fig:model_schematic}
\end{figure}

We establish this phenomenon in a bond-dissipative modified dimerized Kitaev chain, whose Hermitian limit hosts trivial, topological superconducting, and SSH-like regimes \cite{Kitaev2001,PhysRevB.96.205428}. In our case, the bond loss and gain jump operators yield a quadratic Lindblad problem whose Majorana rapidity matrix, at $\mu=0$, separates exactly into the $1\!-\!4$ and $2\!-\!3$ Majorana sectors, labeled by $\eta=\pm$ (see Fig.~\ref{fig:model_schematic}). Each sector forms a non-Hermitian SSH-like rapidity chain with generally distinct effective couplings, GBZs, damping gaps, and non-Bloch windings. The total damping gap does not capture this direct-sum structure, whereas the sector windings predict the number of isolated OBC edge rapidities and establish a sector-resolved non-Bloch bulk-boundary correspondence. For finite \(\mu\), the sectors hybridize necessitating the full rapidity matrix.

Since the Majorana density is sector-blind, resolving the slow channel requires same-sector covariance elements, which are not directly accessible. We therefore turn to the spatial entanglement spectrum (ES), which for a Gaussian state is determined entirely by the covariance matrix restricted to a spatial subsystem \cite{PhysRevLett.101.010504,PhysRevLett.104.130502,PhysRevB.81.064439,PeschelEisler2009}. Within this spectrum, finite-time entanglement {\it{zero events}}, corresponding to crossings or touchings of the entanglement eigenvalues, can diagnose post-quench topology \cite{PhysRevLett.118.185701,PhysRevLett.121.250601,PhysRevLett.121.090401, PhysRevResearch.3.033022}. A recent work has shown that PBC ES dynamics can detect topology defined from the OBC spectrum in a Lindbladian SSH model with a Liouvillian skin effect \cite{PhysRevB.111.L140303}. In contrast, the present superconducting rapidity problem contains two inequivalent Majorana sectors, while physical observables generally mix the sectors at the covariance level. We show that, for balanced gain and loss, finite-time zero events in the total spatial ES signal the presence of nontrivial OBC sector topology even though the physical evolution remains periodic. Together, the sector edge-count rule, the covariance-selection rule, and the entanglement response establish a sector-resolved dynamical correspondence between PBC evolution and hidden OBC topology.

\emph{Model and Majorana-sector rapidity matrix.--}
\label{sec:model_main}
We consider the dimerized Kitaev chain shown in Fig.~\ref{fig:model_schematic}. The coherent Hamiltonian is
\begin{equation}
\begin{aligned}
H={}&-\sum_j\Bigl[
t_1 a_j^\dagger b_j
+\Delta_1 a_j^\dagger b_j^\dagger
+\mathrm{H.c.}\Bigr] \\
&-\sum_j\Bigl[
t_2 b_j^\dagger a_{j+1}
+\Delta_2 b_j^\dagger a_{j+1}^\dagger
+\mathrm{H.c.}\Bigr] \\
&-\mu\sum_j\Bigl[
a_j^\dagger a_j+b_j^\dagger b_j
\Bigr].
\end{aligned}
\label{eq:H_main}
\end{equation}
Here, \(t_1=t(1+d)\), \(t_2=t(1-d)\), \(\Delta_1=\Delta(1-d)\), and \(\Delta_2=\Delta(1+d)\).  In the Hermitian limit, this inverse dimerization produces trivial, topological-superconducting, and SSH-like regimes~\cite{PhysRevB.96.205428}. The bath is described by bond loss and gain jump operators with rates $\gamma_\ell$ and $\gamma_g$~\cite{PhysRevLett.123.170401},

\begin{equation}
    L_j^\ell=\sqrt{\frac{\gamma_\ell}{2}}(a_j-i b_j),
    \qquad
    L_j^g=\sqrt{\frac{\gamma_g}{2}}(a_j^\dagger+i b_j^\dagger),
\label{eq:jumps_main}
\end{equation}
with \(\Gamma=\gamma_\ell+\gamma_g\) and \(\delta\gamma=\gamma_\ell-\gamma_g\). The nonunitary dynamics of the density matrix \(\rho\) is governed by the Lindblad master equation
\begin{equation}
    \dot\rho=-i[H,\rho]+\sum_{j,\alpha=\ell,g}\left(2L_j^\alpha\rho L_j^{\alpha\dagger}-\{L_j^{\alpha\dagger}L_j^\alpha,\rho\}\right).
\end{equation}
Third quantization yields the block-triangular structure matrix
\begin{equation}
    \mathcal A=\begin{pmatrix}-X^\dagger&-iY\\0&X\end{pmatrix}.
\label{eq:structure_main}
\end{equation}
The Liouvillian spectrum is determined by the rapidity matrix \(X\); the source matrix \(Y\) fixes the steady-state covariance but does not modify the rapidity spectrum.  For the local Majorana ordering
\begin{equation}
    W_j=(w^1_{j,A},w^1_{j,B},w^2_{j,A},w^2_{j,B})^T,
\label{eq:ordering_main}
\end{equation}
the Bloch rapidity matrix is
\begin{align}
X(k)={}&
\tfrac{i\Gamma}{2}\tau_0\sigma_0
-\mu\tau_y\sigma_0
-(t_1+t_2\cos k)\tau_y\sigma_x
\nonumber\\
&-\Bigl(t_2\sin k+\tfrac{i\Gamma}{2}\Bigr)\tau_y\sigma_y
+(\Delta_1-\Delta_2\cos k)\tau_x\sigma_y
\nonumber\\
&+\Delta_2\sin k\,\tau_x\sigma_x .
\label{eq:Xk_main}
\end{align}
The derivation of Eq.~\ref{eq:Xk_main} and the finite-chain construction of $X_{\rm OBC}$ are given in Appendix~\ref{sec:rapidity_matrix}, while the resulting PBC/OBC spectra and NHSE diagnostics are presented in Appendix~\ref{sec:rapidity_spectra}. In particular, the OBC spectra are obtained by diagonalizing the \(4N\times4N\) block-tridiagonal rapidity matrix \(X_{\rm OBC}\), constructed from the real-space blocks of \(X(k)\).
At \(\mu=0\), \(X(k)\) couples only the pairs \((1,4)\) and \((2,3)\) in the ordering of Eq.~\eqref{eq:ordering_main}, as illustrated in Fig.~\ref{fig:model_schematic}(b). Thus the rapidity problem separates into two independent Majorana sectors,
\begin{equation}
    \text{1--4} : (w^1_A,w^2_B), \qquad \text{2--3} : (w^1_B,w^2_A),
\label{eq:sector_main}
\end{equation}
which we denote by \(\eta=+\) and \(\eta=-\), respectfully. The corresponding effective intracell
and intercell couplings are
\begin{equation}
    a_\eta=t_1-\eta\Delta_1,
    \qquad
    b_\eta=t_2+\eta\Delta_2,
    \qquad \eta=\pm .
\label{eq:ab_main}
\end{equation}
Each sector is therefore a non-Hermitian SSH-like Majorana chain, as shown in Fig.~\ref{fig:model_schematic}(c). The mapping is exact, with $a_\eta$ and $b_\eta$ playing the roles of the SSH intracell and intercell hoppings $t_1$ and $t_2$, while the bath rate $\Gamma$ enters only as a nonreciprocity of the intracell bond, whose amplitudes are $a_\eta\pm\eta\Gamma/2$ along the two directions. Each sector is thus a non-Hermitian SSH chain \cite{PhysRevLett.121.086803}. This sector decomposition is exact only at \(\mu=0\). For finite $\mu$, the two Majorana sectors hybridize, so the sector GBZs and windings are no longer separately defined; the corresponding full-matrix crossover is analyzed
in Appendix~\ref{secS:finite_mu}.

\emph{Sector non-Bloch topology and edge rapidities.---}
\label{sec:topology_main}
\begin{figure}[t]
    \centering
    \includegraphics[width=\columnwidth]{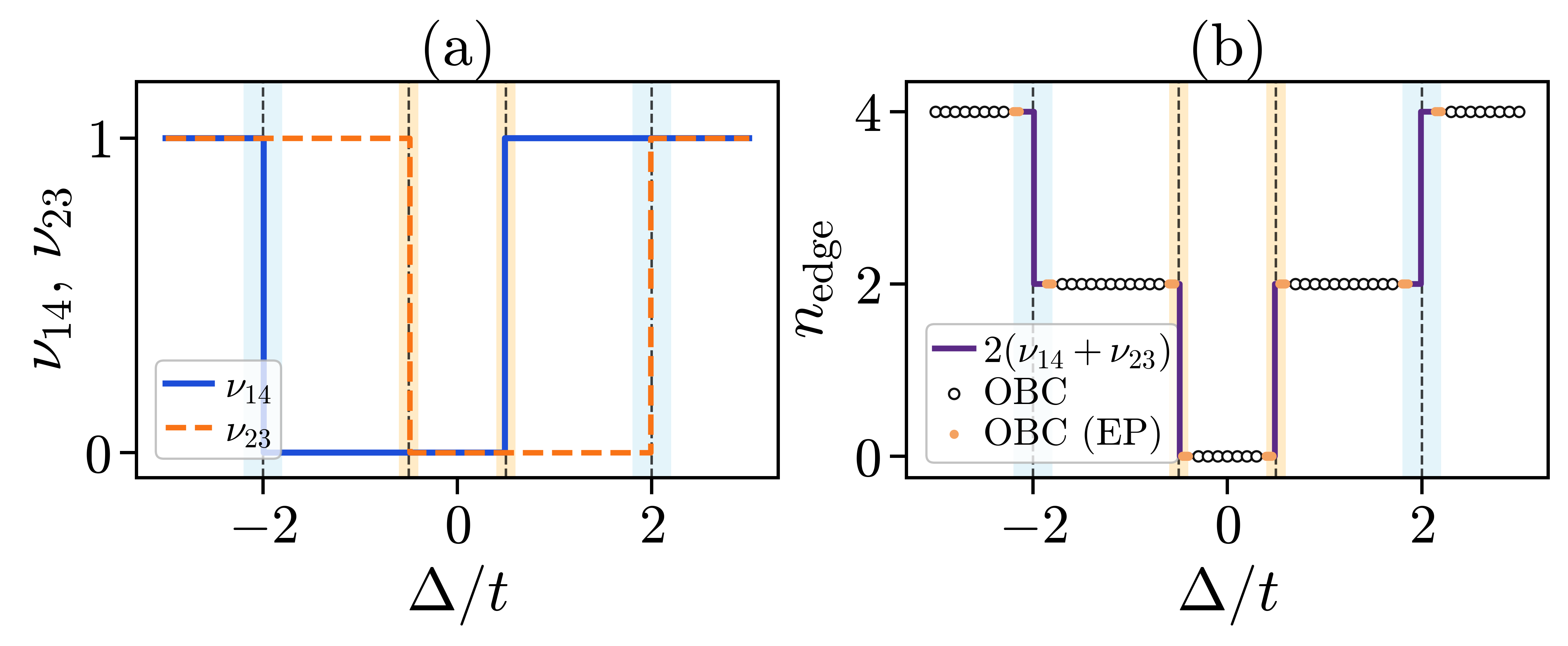}
    \caption{
    (a) Non-Bloch winding numbers \(\nu_{14}\) and \(\nu_{23}\) as functions of \(\Delta/t\). 
    (b) Number of isolated edge rapidities in the finite OBC rapidity spectrum. The purple line shows the sector-resolved prediction \(2(\nu_{14}+\nu_{23})\) from Eq.~\eqref{eq:edge_count_main}, while symbols are obtained by direct diagonalization of \(X_{\rm OBC}\). Open circles denote OBC counts away from the shaded real-gapless windows, and filled orange circles denote counts evaluated inside those windows. The shaded regions mark the PBC real-gapless windows bounded by exceptional points, and the vertical dashed lines indicate the Hermitian critical values \(\Delta/t=\pm d\) and \(\Delta/t=\pm1/d\). Parameters are \(t=1\), \(d=0.5\), \(\mu=0\), and \(\gamma_\ell=\gamma_g=0.2\).
    }
    \label{fig:sector_topology}
\end{figure}
\begin{figure*}[t]
    \centering
    \includegraphics[width=0.92\textwidth]{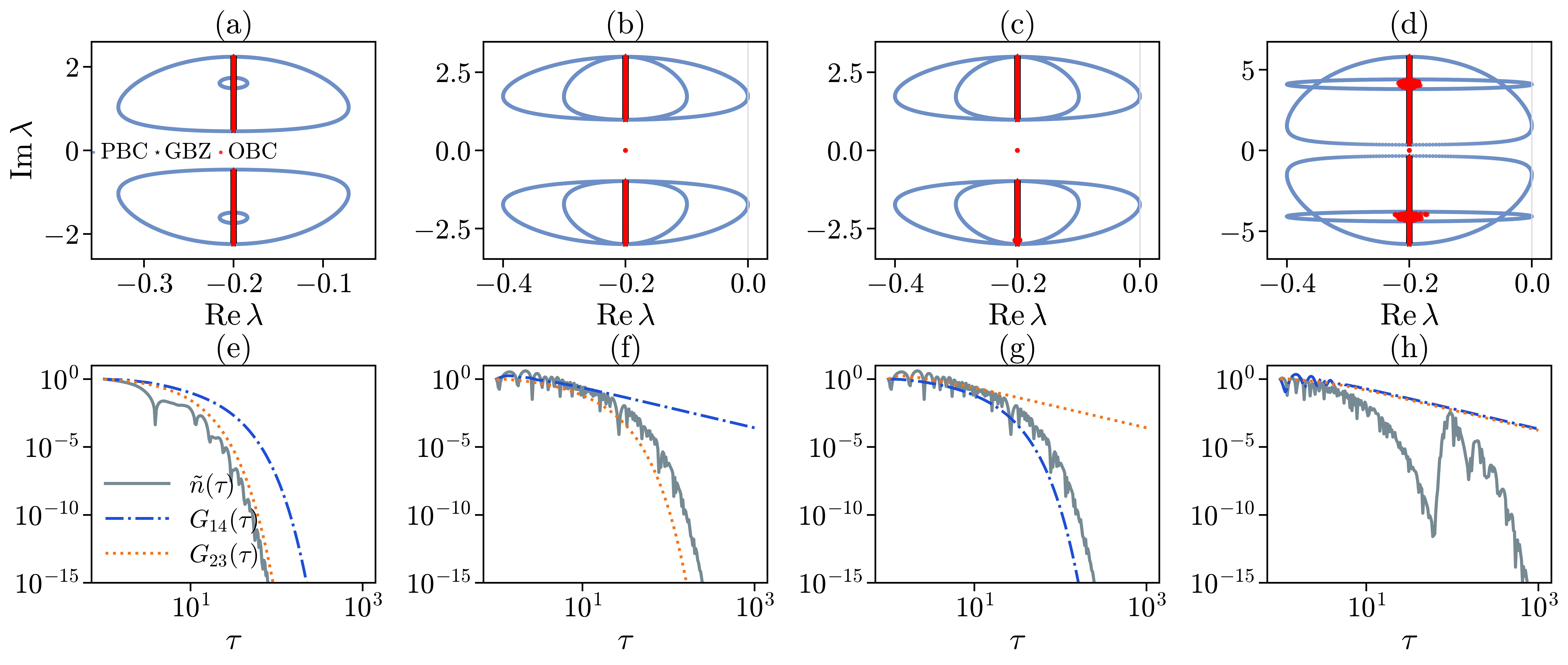}
    \caption{
Sector-resolved rapidity spectra and PBC damping.
(a)--(d) Damping-plane spectra \(\lambda=i\beta\), showing PBC Bloch bands, sector-GBZ continua, and finite-chain OBC rapidities. (e)--(h) Normalized root-mean-square amplitudes \(\widetilde n(\tau)\), \(G_{14}(\tau)\), and \(G_{23}(\tau)\). Columns correspond to \(\Delta/t=0.25,1,-1,\) and \(2.5\), representing the trivial, \(1_{ab}\), \(1_{ba}\), and SSH-like regimes. Only probes overlapping a gapless PBC sector show algebraic damping; the cross-sector density need not. Parameters are \(t=1\), \(d=0.5\), \(\mu=0\), and \(\gamma_\ell=\gamma_g=0.2\).
}
    \label{fig:damping}
\end{figure*}
Before defining the sector invariants, we first note why a non-Bloch description is required for the present model. Direct diagonalization of the Bloch rapidity matrix \(X(k)\) under PBC and \(X_{\rm OBC}\) under OBC gives qualitatively different spectra. Finite bond dissipation also broadens the Hermitian bulk gap closings into finite real-gapless rapidity windows in the PBC spectrum. The boundaries of these windows are second-order exceptional points identified by the vanishing of the biorthogonal phase rigidity. The complete PBC/OBC spectra, the right-eigenvector
accumulation demonstrating the NHSE, and the
exceptional-point diagnostics are presented in Appendix~\ref{sec:rapidity_spectra}. The key point for the following discussion is that the OBC bulk rapidity continuum is not obtained from the Bloch unit circle \(z=e^{ik}\). Instead, it is obtained from a sector-dependent non-Bloch continuation.

Under OBC, each Majorana sector has its own GBZ,
\begin{equation}
    z=r_\eta e^{ik},\qquad
    r_\eta=\sqrt{\left|\frac{a_\eta-\Gamma/2}{a_\eta+\Gamma/2}\right|},
\label{eq:gbz_main}
\end{equation}
and the bulk rapidity continuum is the union of the two sector-resolved non-Bloch continua. The sector-dependent GBZ construction and the resulting non-Bloch bulk continua are derived in Appendix~\ref{sec:sector_nonbloch}. Excluding the uniform shift $i\Gamma/2$, each sector block is purely off-diagonal and therefore chiral symmetric, which quantizes the non-Bloch winding~\cite{PhysRevLett.124.040401}. The sector winding number then reduces to
\begin{equation}
\nu_\eta=
\begin{cases}
1,& |b_\eta|^2>|a_\eta^2-\Gamma^2/4|,\\
0,& |b_\eta|^2<|a_\eta^2-\Gamma^2/4|,
\end{cases}
\qquad \eta=\pm .
\label{eq:winding_main}
\end{equation}
At equality, the non-Bloch point gap closes at the shifted rapidity \(\beta=i\Gamma/2\). A nonzero sector winding contributes two isolated OBC edge rapidities, so that
\begin{equation}
    n^{\rm pred}_{\rm edge}=2(\nu_{14}+\nu_{23}).
\label{eq:edge_count_main}
\end{equation}
The non-Bloch winding criterion and the additive
edge-count rule are derived in Appendix~\ref{sec:sector_nonbloch}. Following the convention for the modified dimerized Kitaev chain, we refer to the regions with \((\nu_{14},\nu_{23})=(1,0)\) and \((0,1)\) (see Fig.~\ref{fig:sector_topology}) as the \(1_{ab}\) and \(1_{ba}\) topological-superconductor regimes, respectively, while \((1,1)\) denotes the SSH-like regime and \((0,0)\) the trivial regime.

Figure~\ref{fig:sector_topology} verifies this sector-resolved non-Bloch bulk-boundary correspondence. In the \(1_{ab}\) region only the $1\!-\!4$ sector is topological; in the \(1_{ba}\) region only the $2\!-\!3$ sector is topological; and in the SSH-like region both sectors are topological. This sector resolution is crucial for the dynamical analysis, because an observable can project onto sector $1\!-\!4$, sector $2\!-\!3$, or a cross-sector channel.

\emph{Observable-selective damping.---}
\label{sec:damping_main}
We next show that a gapless sector need not dominate every local observable. We use the damping convention \(\lambda_n=i\beta_n\), so stable modes have \({\rm Re}\,\lambda_n\le0\). Following the damping-matrix analysis of the dissipative SSH chain \cite{PhysRevLett.123.170401}, we define
\begin{equation}
    \Lambda_{\rm PBC}=\min_{n,k}\left[-2\,{\rm Re}\,\lambda_n(k)\right].
\label{eq:liouvillian_gap_main}
\end{equation}
A finite \(\Lambda_{\rm PBC}\) gives exponential convergence, while \(\Lambda_{\rm PBC}=0\) permits algebraic relaxation in the thermodynamic limit if the observable has nonzero projection onto the gapless damping channel. At \(\mu=0\), \(X(k)\) separates into the $1\!-\!4$ and $2\!-\!3$ sectors. Thus, we define
\begin{equation}
    \Lambda_{\eta}=\min_{\alpha=\pm,k}\left[-2\,{\rm Re}\,\lambda_{\eta,\alpha}(k)\right],
    \qquad \eta=\pm .
\label{eq:sector_gap_main}
\end{equation}
For \(a_\eta=t_1-\eta\Delta_1\) and \(b_\eta=t_2+\eta\Delta_2\), the PBC sector gap vanishes when \(|a_\eta|\le |b_\eta|\), the Majorana-sector analogue of the dissipative SSH gapless condition.

For the vector \(W=(W_1,\ldots,W_{4N})^T\), denoting the Majoranas, we define the covariance matrix
\begin{equation}
    C_{mn}(\tau)=\frac{i}{2}{\rm Tr}\{\rho(\tau)[W_m,W_n]\}
    =i\left(\langle W_mW_n\rangle_\tau-\delta_{mn}\right).
\label{eq:cov_def_main}
\end{equation}
Each element \(C_{mn}\) correlates two Majorana operators, \(W_m\) and \(W_n\). For balanced gain and loss, \(\gamma_\ell=\gamma_g\), the source term vanishes and the covariance deviation \(\widetilde C(\tau)=C(\tau)-C_{\rm ss}\) evolves as
\begin{equation}
    \widetilde{C}(\tau) = e^{(iX)^T \tau} \widetilde{C}(0) e^{iX\tau}.
\label{eq:cov_dynamics_main}
\end{equation}
Since the evolution operator acts from both left and right, a covariance eigenmode decays with a sum of two damping eigenvalues. The rapidity spectrum therefore identifies the slow sectors, while the observable determines which covariance entries are actually measured.

The physical density is not a same-sector covariance. Using,
\begin{equation}
    n_{j,A}=\frac{1+i w^1_{j,A}w^2_{j,A}}{2},
    \qquad
    n_{j,B}=\frac{1+i w^1_{j,B}w^2_{j,B}}{2},
\end{equation}
the unit-cell density deviation is
\begin{equation}
    \widetilde n_j(\tau)=\frac{1}{2}
    \left[\widetilde C_{jA1,jA2}(\tau)+\widetilde C_{jB1,jB2}(\tau)\right],
\label{eq:density_main}
\end{equation}
where $jA1$ denotes $w^1_{j,A}$, etc. Both terms in Eq.~(17) connect the $1\!-\!4$ and $2\!-\!3$ sectors,
so the density decay involves cross-sector sums of damping
eigenvalues. Consequently, when only one sector is
gapless, the other can keep the relevant density channel
exponentially damped. For the relaxation curves in
Fig.~\ref{fig:damping}, we define the normalized
root-mean-square amplitude
\begin{equation}
\mathcal{R}_{f}(\tau)=
\frac{
\left[N^{-1}\sum_j |f_j(\tau)|^2\right]^{1/2}
}{
\left[N^{-1}\sum_j |f_j(\tau_{\rm ref})|^2\right]^{1/2}
},
\qquad \tau_{\rm ref}=1,
\label{eq:rms_amplitude}
\end{equation}
where $\tau_{\rm ref}$ is the earliest time shown in
Fig.~\ref{fig:damping}. Taking
$f_j=\widetilde n_j$, $\widetilde C_{jA1,jB2}$, and
$\widetilde C_{jB1,jA2}$ defines
$\widetilde n(\tau)$, $G_{14}(\tau)$, and
$G_{23}(\tau)$, respectively. The latter two are auxiliary
same-sector covariance probes and are not physical
densities.

\begin{figure*}[t]
    \centering
    \includegraphics[width=0.98\textwidth]{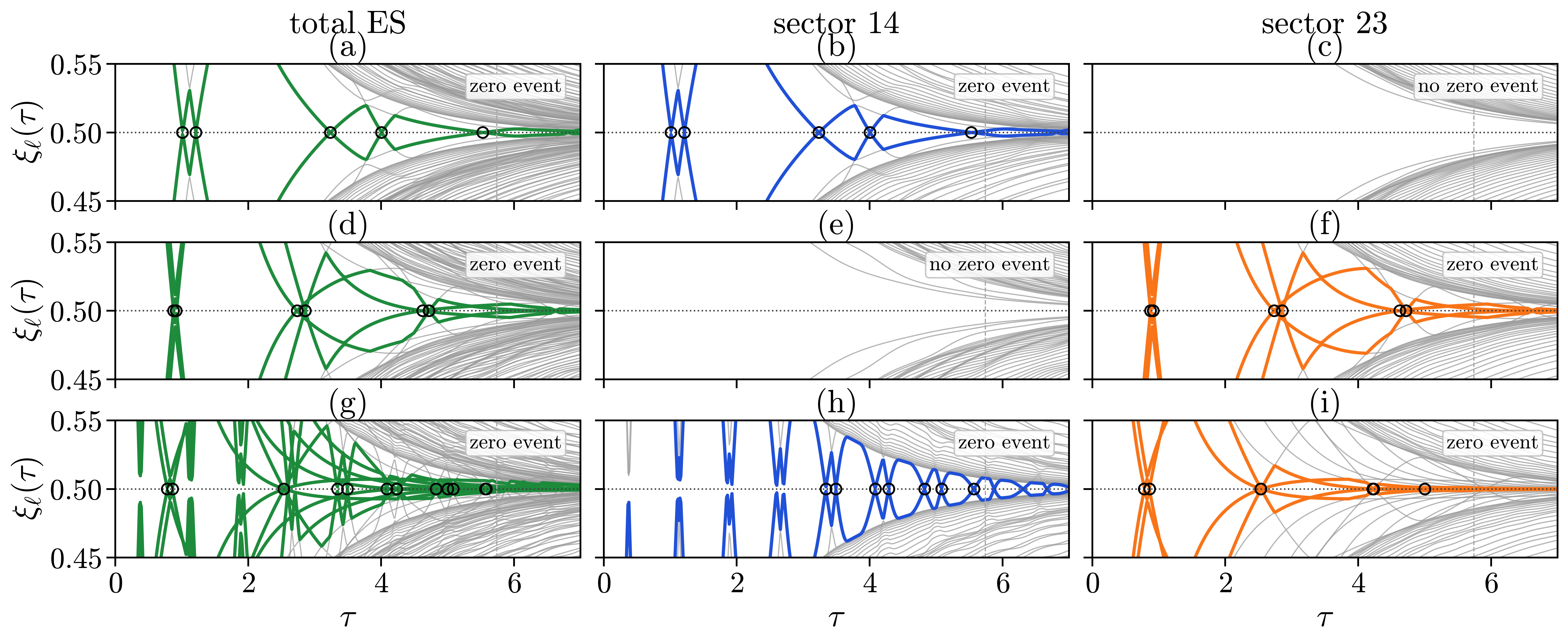}
    \caption{
    Entanglement-spectrum dynamics under PBC after dissipative quenches from a trivial Hermitian initial state at \(\Delta_i/t=0.25\). Rows correspond to post-quench values \(\Delta/t=1,-1,\) and \(2.5\), with sector windings \((\nu_{14},\nu_{23})=(1,0),(0,1),\) and \((1,1)\). Columns show the total ES and the sector-restricted covariance ES diagnostics for sectors $1\!-\!4$ and $2\!-\!3$. Plotted are the entanglement occupations \(\xi_\ell(\tau)\), with \(\xi_\ell=1/2\) marking an entanglement zero mode. Colored curves highlight branches participating in finite-time zero events, gray curves represent the remaining branches, and open circles mark zero events. Parameters are \(t=1\), \(d=0.5\), \(\mu=0\), \(\gamma_\ell=\gamma_g=0.2\), \(N=200\), and \(N_S=100\).
    }
    \label{fig:ES}
\end{figure*}

The upper row of Fig.~\ref{fig:damping} identifies the slow channels. In the trivial regime, both sectors are PBC-gapped. In the \(1_{ab}\) regime the gapless PBC channel belongs to sector $1\!-\!4$, whereas in the \(1_{ba}\) regime it belongs to sector $2\!-\!3$. Also, in the SSH-like regime both the sector gaps vanish. The GBZ curves and OBC points show the corresponding non-Bloch open-chain spectra, whose complete spectral and topological constructions are given in Appendices~\ref{sec:rapidity_spectra} and \ref{sec:sector_nonbloch}. The lower row demonstrates the observable selection rule. In the trivial regime, all traces decay rapidly. In the \(1_{ab}\) regime, \(G_{14}\) shows slow algebraic damping, while \(G_{23}\) and the physical density decay faster. In the \(1_{ba}\) regime, the roles are reversed and the slow channel is \(G_{23}\). In the SSH-like regime, both same-sector probes are slow. Thus, a sector gap closing produces algebraic relaxation only for covariance perturbations that overlap with that sector. The physical density, being a cross-sector covariance, need not inherit the algebraic decay of a single gapless Majorana sector.

\emph{Entanglement-spectrum dynamics.---}
\label{sec:es_main}
The damping analysis shows that local observables do not always provide a complete diagnostic of the sector topology. We therefore consider the ES of a spatial subsystem, following the zero-event diagnostic of Ref.~\cite{PhysRevB.111.L140303}. The protocol is a dissipative quench for which the initial covariance matrix \(C_i\equiv C(0)\) is prepared as the ground-state covariance of the isolated Hermitian chain at \(t=1\), \(d=0.5\), \(\mu=0\), and \(\Delta_i/t=0.25\), with no bath present. At \(\tau=0^+\), the pairing is changed to the post-quench value \(\Delta\), where balanced bond dissipation with \(\gamma_\ell=\gamma_g=0.2\) is turned on, and the covariance matrix evolves according to the corresponding PBC rapidity matrix.

We bipartition the periodic chain into two subsystems \(S\) each with \(N_S\) consecutive unit cells. In Fig.~\ref{fig:ES}, \(N=200\) and \(N_S=100\). The physical chain is not opened and the cut is solely used to form the subsystem covariance matrix $C_S(\tau)$ from the full PBC covariance matrix $C(\tau)$. Unlike the local density, $C_S$ retains all the Majorana pairs inside $S$, including the same-sector blocks, and therefore has access to the slow channels, for which the density is blind. If $\zeta_\ell(\tau)$ denotes the eigenvalues of $iC_S(\tau)$, we plot

\begin{equation}
    \xi_\ell(\tau)=\frac{1+\zeta_\ell(\tau)}{2},
\label{eq:xi_main}
\end{equation} as a function of time $\tau$.
An entanglement zero mode corresponds to \(\xi_\ell=1/2\). For balanced gain and loss, the source term in the covariance equation vanishes and \(C_{\rm ss}=0\), so all \(\xi_\ell\) approach a value \(1/2\) at long times. We therefore count only finite-time crossings or intersection of \(\xi_\ell=1/2\), before the final asymptotic collapse occurs.

We compute the total ES using all the Majoranas in \(S\). At \(\mu=0\), the rapidity matrix block-diagonalizes into the independent \(1\!-\!4=(w_A^1,w_B^2)\) and \(2\!-\!3=(w_B^1,w_A^2)\) sectors. We, therefore also compute diagnostic sector-restricted covariance spectra by retaining only the Majorana components of one sector inside \(S\). For the quench to \(\Delta/t=1\), zero events appear in the total ES and sector $1\!-\!4$ ES, but not in sector $2\!-\!3$. For \(\Delta/t=-1\), the behavior is reversed. For \(\Delta/t=2.5\), both sector-restricted spectra show zero events. Thus, finite-time ES zero events occur precisely in the sectors whose post-quench OBC/non-Bloch winding is nonzero. The physical evolution remains periodic throughout; the ES cut reveals the sector-resolved open-boundary topology encoded in the rapidity matrix. The entanglement dynamics, thus functions as a dynamical topological invariant, a nonlocal quantity evaluated under PBC that returns the OBC edge content. The trivial-to-trivial control and systematic zero-event scan are given in Appendix~\ref{secS:ES_scan}, while the finite-$\mu$ full-matrix check is presented in Appendix~\ref{secS:finite_mu}.

\emph{Conclusion.---}
We have shown that the correspondence between the Liouvillian damping gap and observable relaxation, earlier established for the dissipative SSH chain, breaks down in a dissipative topological superconductor. In a bond-dissipative dimerized Kitaev chain at \(\mu=0\), the Majorana rapidity matrix separates into two sectors, each with its own GBZ and winding number, and the sector windings determine the number of isolated open-boundary edge rapidities, establishing a sector-resolved non-Bloch bulk-boundary correspondence. The same decomposition makes the relaxation observables dependent, since the local density is a cross-sector covariance and it decays at the sum of the two sector rates, so that it remains exponentially damped for the gapless sector, whereas same-sector covariance probes resolve the slow channel. For balanced gain and loss, the ES dynamics under periodic-boundary Lindblad evolution detect this hidden open-boundary topology through finite-time zero events, providing a nonlocal probe that requires no physical breaking of the chain.

\begin{acknowledgments}
S.G. acknowledges the Anusandhan National Research Foundation (ANRF), Govt. of India, for providing financial support through the National Post Doctoral Fellowship (NPDF) (File No. PDF/2025/004365). KR acknowledges the research
fellowship from the MoE, Government of India. BT acknowledges support from the Scientific and Technological Research Council of Türkiy (TÜBİTAK) under Grant No.~125F435 and from the Turkish Academy of Sciences (TÜBA) under Grant No ~AD-2026. SB acknowledges support from TÜBİTAK-BİDEB.

\end{acknowledgments}

\onecolumngrid
\appendix

\makeatletter
\@addtoreset{figure}{section}
\makeatother
\renewcommand{\thefigure}{\Alph{section}\arabic{figure}}

\section{Rapidity matrix and finite-chain construction}
\label{sec:rapidity_matrix}

We first derive the Majorana rapidity matrix used in the main text. We use the local Majorana ordering
\[
    W_j=
    \left(
    w^1_{j,A},\,
    w^1_{j,B},\,
    w^2_{j,A},\,
    w^2_{j,B}
    \right)^T .
\]
In this basis, the Pauli matrices \(\tau_\alpha\) act on the
Majorana flavor index \((1,2)\), while \(\sigma_\alpha\) act on the sublattice index \((A,B)\). The Lindblad equation in the main text is written as
\begin{equation}
    \dot{\rho}
    =
    -i[H,\rho]
    +
    \sum_\alpha
    \left(
    2L_\alpha\rho L_\alpha^\dagger
    -
    \{L_\alpha^\dagger L_\alpha,\rho\}
    \right).
\label{eq:app_lindblad_convention}
\end{equation}
With this convention, the even-parity third-quantized structure matrix has the block-triangular form \cite{Prosen2008,Prosen2010}
\begin{equation}
    \mathcal A
    =
    \begin{pmatrix}
        -X^\dagger & -iY\\
        0 & X
    \end{pmatrix}.
\label{eq:app_structure_matrix}
\end{equation}
Here
\[
    X=4H_M+i(M+M^T),
    \qquad
    Y=2(M-M^T).
\]
Here \(H_M\) is the closed-system Majorana matrix and \(M\) is the bath matrix constructed from the linear jump operators. The eigenvalues of \(X\) are the rapidities. We also define
\[
    \Gamma=\gamma_\ell+\gamma_g,
    \qquad
    \delta\gamma=\gamma_\ell-\gamma_g .
\]
Below we show that only the total rate \(\Gamma\) enters \(X\),
whereas the imbalance \(\delta\gamma\) enters only the source block
\(Y\).

Using
\[
\begin{alignedat}{2}
    w^1_{j,A} &= a_j+a_j^\dagger,
    \qquad&
    w^2_{j,A} &= -i(a_j-a_j^\dagger),
    \\
    w^1_{j,B} &= b_j+b_j^\dagger,
    \qquad&
    w^2_{j,B} &= -i(b_j-b_j^\dagger),
\end{alignedat}
\]
or equivalently
\[
    a_j=\frac{w^1_{j,A}+i w^2_{j,A}}{2},
    \qquad
    b_j=\frac{w^1_{j,B}+i w^2_{j,B}}{2},
\]
the coherent Hamiltonian gives
\begin{align}
4H_M(k)
=&
-\mu\,\tau_y\sigma_0
-
\left(t_1+t_2\cos k\right)\tau_y\sigma_x
-
t_2\sin k\,\tau_y\sigma_y
\nonumber\\
&+
\left(\Delta_1-\Delta_2\cos k\right)\tau_x\sigma_y
+
\Delta_2\sin k\,\tau_x\sigma_x .
\label{eq:app_4HM}
\end{align}
The intracell coherent bond loss and gain jump operators are
\[
    L_j^\ell
    =
    \sqrt{\frac{\gamma_\ell}{2}}
    \left(a_j-i b_j\right),
    \qquad
    L_j^g
    =
    \sqrt{\frac{\gamma_g}{2}}
    \left(a_j^\dagger+i b_j^\dagger\right).
\]
In the above Majorana basis,
\[
    L_j^\ell=(l_j^\ell)^T W_j,
    \qquad
    L_j^g=(l_j^g)^T W_j,
\]
with
\begin{equation}
    l_j^\ell
    =
    \sqrt{\frac{\gamma_\ell}{8}}
    \begin{pmatrix}
    1\\ -i\\ i\\ 1
    \end{pmatrix},
    \qquad
    l_j^g
    =
    \sqrt{\frac{\gamma_g}{8}}
    \begin{pmatrix}
    1\\ i\\ -i\\ 1
    \end{pmatrix}.
\label{eq:app_l_vectors}
\end{equation}
Owing to the convention in Eq.~\eqref{eq:app_lindblad_convention},
the local bath matrix is
\[
    M_j
    =
    2\left(
    l_j^\ell l_j^{\ell\dagger}
    +
    l_j^g l_j^{g\dagger}
    \right).
\]
Substituting Eq.~\eqref{eq:app_l_vectors}, the symmetric and
antisymmetric combinations are
\begin{align}
    M_j+M_j^T
    &=
    \frac{\Gamma}{2}
    \left(
    \tau_0\sigma_0-\tau_y\sigma_y
    \right),
\label{eq:app_M_plus_MT_pauli}
\\
    Y_j=2(M_j-M_j^T)
    &=
    \delta\gamma
    \left(
    \tau_y\sigma_0-\tau_0\sigma_y
    \right).
\label{eq:app_Y_pauli}
\end{align}
Therefore, the dissipative contribution to the rapidity matrix is
\[
    X_{\rm diss}
    =
    i(M_j+M_j^T)
    =
    \frac{i\Gamma}{2}
    \left(
    \tau_0\sigma_0-\tau_y\sigma_y
    \right).
\]
The gain-loss imbalance \(\delta\gamma\) appears only in \(Y\), not in
\(X\). Hence the rapidity spectrum is controlled by
\(\Gamma=\gamma_\ell+\gamma_g\).
Combining the coherent contribution in Eq.~\eqref{eq:app_4HM} with the dissipative contribution, we obtain
\begin{equation}
\begin{aligned}
X(k)
=&\;
\frac{i\Gamma}{2}\tau_0\sigma_0
-\mu\,\tau_y\sigma_0
-\left(t_1+t_2\cos k\right)\tau_y\sigma_x
\\
&-
\left(t_2\sin k+\frac{i\Gamma}{2}\right)\tau_y\sigma_y
+
\left(\Delta_1-\Delta_2\cos k\right)\tau_x\sigma_y
\\
&+
\Delta_2\sin k\,\tau_x\sigma_x .
\end{aligned}
\label{eq:Xk}
\end{equation}
This is the Bloch rapidity matrix used in the main text. The term
\(-(i\Gamma/2)\tau_y\sigma_y\) originates from the coherent bond
gain-loss dissipators, while the uniform term
\((i\Gamma/2)\tau_0\sigma_0\) gives the overall dissipative shift of
the rapidity spectrum.

For finite chains, we write
\begin{equation}
    X(k)=X_0+X_{+1}e^{ik}+X_{-1}e^{-ik}.
\label{eq:X_blocks}
\end{equation}
The open-boundary rapidity matrix is the block-tridiagonal matrix
\begin{equation}
X_{\rm OBC}
=
\begin{pmatrix}
X_0      & X_{+1}  & 0       & \cdots & 0\\
X_{-1}  & X_0     & X_{+1}  & \cdots & 0\\
0       & X_{-1} & X_0      & \ddots & \vdots\\
\vdots  & \vdots & \ddots   & \ddots & X_{+1}\\
0       & 0      & \cdots   & X_{-1} & X_0
\end{pmatrix}.
\label{eq:X_OBC}
\end{equation}
The periodic finite-chain matrix is obtained by adding the corner blocks \(X_{-1}\) at the upper-right corner and \(X_{+1}\) at the lower-left corner. Explicitly, with \(\sigma_\pm=(\sigma_x\pm i\sigma_y)/2\),
\begin{align}
X_0
=&
-\mu\,\tau_y\sigma_0
-t_1\,\tau_y\sigma_x
+\Delta_1\,\tau_x\sigma_y
+
\frac{i\Gamma}{2}
\left(
\tau_0\sigma_0-\tau_y\sigma_y
\right),
\label{eq:X0_block}
\\
X_{+1}
=&
-\left(t_2\tau_y+i\Delta_2\tau_x\right)\sigma_-,
\label{eq:Xplus_block}
\\
X_{-1}
=&
-\left(t_2\tau_y-i\Delta_2\tau_x\right)\sigma_+ .
\label{eq:Xminus_block}
\end{align}
Unless stated otherwise, all finite-chain OBC spectra and eigenvectors are obtained by diagonalizing Eq.~\eqref{eq:X_OBC}.

At \(\mu=0\), the rapidity matrix separates into two Majorana sectors,
\begin{equation}
    \eta=+:(w^1_A,w^2_B),
    \qquad
    \eta=-:(w^1_B,w^2_A),
\label{eq:eta_sector_definition}
\end{equation}
with effective couplings
\begin{equation}
    a_\eta=t_1-\eta\Delta_1,
    \qquad
    b_\eta=t_2+\eta\Delta_2,
    \qquad \eta=\pm .
\label{eq:aeta_beta}
\end{equation}
The corresponding PBC rapidities are
\begin{equation}
    \beta^{\rm PBC}_{\eta,\pm}(k)=\frac{i\Gamma}{2}\pm\sqrt{R_\eta(e^{ik})},
\label{eq:beta_pbc}
\end{equation}
where
\begin{equation}
    R_\eta(z)=
    \left(a_\eta+\frac{\Gamma}{2}+b_\eta z^{-1}\right)
    \left(a_\eta-\frac{\Gamma}{2}+b_\eta z\right).
\label{eq:Reta_z}
\end{equation}
The sector-dependent GBZs and the non-Bloch winding criterion are derived in Sec.~\ref{sec:sector_nonbloch}.

\section{Rapidity spectra, exceptional windows, and NHSE}
\label{sec:rapidity_spectra}

\subsection{PBC and OBC rapidity spectra}
\label{subsec:pbc_obc_spectra}
\begin{figure}[t]
    \centering
    \includegraphics[width=0.95\columnwidth]{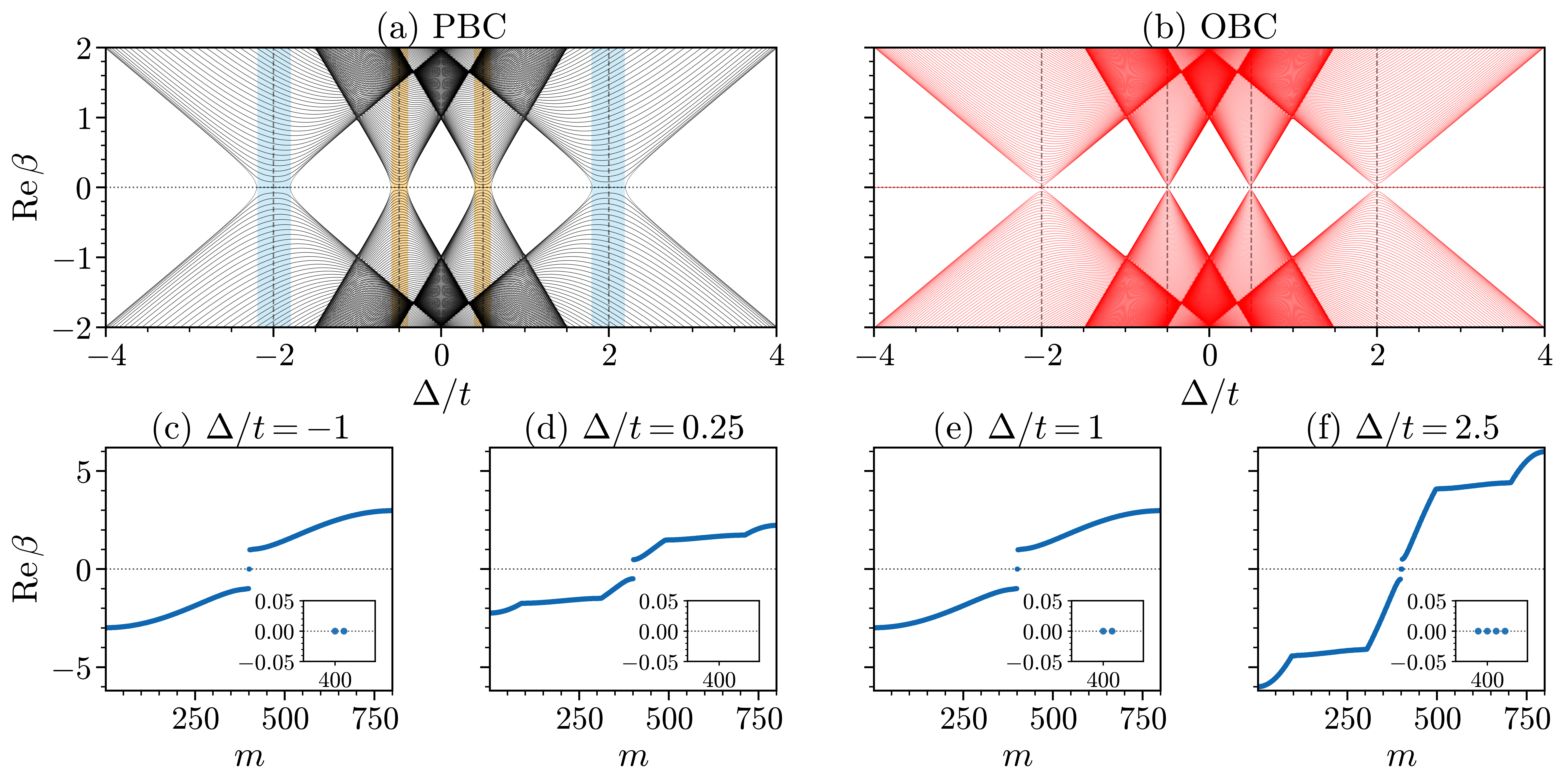}
    \caption{
    PBC and OBC rapidity spectra of the bond-dissipative dimerized Kitaev chain. (a) PBC spectrum obtained by diagonalizing \(X(k)\). (b) OBC spectrum obtained by diagonalizing the finite matrix \(X_{\rm OBC}\). In panels (a) and (b), we plot \(\mathrm{Re}\,\beta\) as a function of \(\Delta/t\). The horizontal dotted line marks \(\mathrm{Re}\,\beta=0\). The vertical dashed lines indicate the Hermitian critical values \(\Delta/t=\pm d\) and \(\Delta/t=\pm1/d\). The shaded regions mark the finite real-gapless windows discussed in Sec.~\ref{subsec:exceptional_windows}. (c)--(f) Sorted OBC rapidities \(\mathrm{Re}\,\beta_m\) at \(\Delta/t=-1,0.25,1,2.5\), respectively. The insets zoom near \(\mathrm{Re}\,\beta=0\), where isolated edge rapidities appear in the nontrivial sectors. Parameters are \(t=1\), \(d=0.5\), \(\mu=0\), and \(\gamma_\ell=\gamma_g=0.2\), so that \(\Gamma=0.4\).
    }
\label{fig:pbc_obc_rapidity_spectra}
\end{figure}
We next compare the rapidity spectra under periodic and open boundary conditions. The single-particle rapidities are the eigenvalues of the Majorana rapidity matrix \(X\). We denote them by \(\beta_n\) and write the corresponding right-eigenvalue problem as
\begin{equation}
    X u_n^R = \beta_n u_n^R .
\label{eq:rapidity_eigenproblem}
\end{equation}
For PBC, we diagonalize the Bloch rapidity matrix \(X(k)\) in Eq.~\eqref{eq:Xk}. At \(\mu=0\), this is equivalently the union of the two sector spectra in Eq.~\eqref{eq:beta_pbc}. For OBC, we diagonalize the finite matrix \(X_{\rm OBC}\) in Eq.~\eqref{eq:X_OBC}.

Figure~\ref{fig:pbc_obc_rapidity_spectra} shows two distinct consequences of bond dissipation. First, in the closed modified dimerized Kitaev chain, the Hermitian bulk gap closes at
\begin{equation}
    \frac{\Delta}{t}=\pm d,
    \qquad
    \frac{\Delta}{t}=\pm\frac{1}{d},
\label{eq:hermitian_critical_points}
\end{equation}
which separate the trivial, topological-superconductor-like, and SSH-like regimes \cite{PhysRevB.96.205428}. For the parameters in Fig.~\ref{fig:pbc_obc_rapidity_spectra}, these critical values are \(\Delta/t=\pm0.5\) and \(\Delta/t=\pm2\). In the dissipative problem, finite bond loss and gain do not shift these critical centers; instead, each Hermitian gap closing broadens into a finite real-gapless window in the PBC rapidity spectrum. Second, the OBC spectrum differs qualitatively from the PBC spectrum. This boundary sensitivity is the spectral signature of non-Bloch physics: the OBC bulk rapidity spectrum is not obtained from the Bloch unit circle, but from the sector-dependent GBZs derived in Sec.~\ref{sec:sector_nonbloch}. Thus the open-chain bulk spectrum is the union of two sector-resolved non-Bloch continua, together with isolated edge rapidities when the sector windings are nonzero.

Panels~\ref{fig:pbc_obc_rapidity_spectra}(c)--\ref{fig:pbc_obc_rapidity_spectra}(f) show the finite-chain OBC spectra at fixed values of \(\Delta/t\). In these panels, the real parts of the OBC rapidities are sorted and plotted as \(\mathrm{Re}\,\beta_m\) versus the mode index \(m\), while the insets zoom near \(\mathrm{Re}\,\beta=0\). We observe no isolated rapidity near \(\mathrm{Re}\,\beta=0\) at the trivial point \(\Delta/t=0.25\). At \(\Delta/t=-1\) and \(\Delta/t=1\), two isolated rapidities appear near \(\mathrm{Re}\,\beta=0\), indicating that one Majorana sector is nontrivial. At the SSH-like point \(\Delta/t=2.5\), four isolated rapidities are visible, indicating that both Majorana sectors are nontrivial. This observed sequence, \(0,2,2,4\), is explained by the sector non-Bloch edge-count rule derived in Sec.~\ref{sec:sector_nonbloch}.

\subsection{Right-eigenvector profiles and NHSE}
\label{subsec:nhse_profiles}
\begin{figure}[t]
    \centering
    \includegraphics[width=0.85\columnwidth]{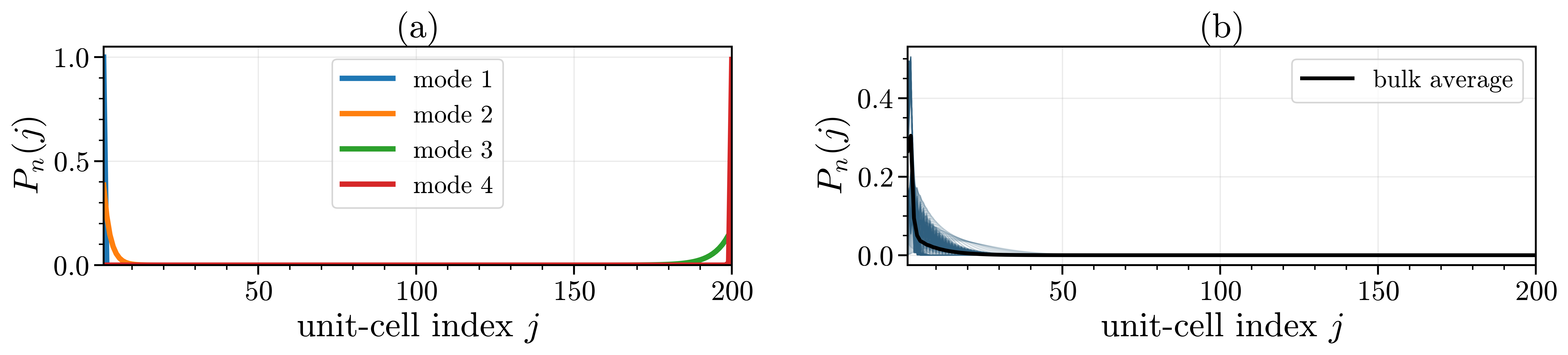}
\caption{
Right-eigenvector profiles of the finite OBC rapidity matrix at \(\Delta/t=2.5\). For each normalized right eigenvector \(X_{\rm OBC}u_n^R=\beta_n u_n^R\), we plot the unit-cell-resolved weight \(P_n(j)\) defined in Eq.~\eqref{eq:cell_resolved_weight}. (a) Profiles of the four isolated edge rapidities closest to the shifted zero rapidity \(\beta=i\Gamma/2\). (b) Profiles of the remaining OBC right eigenvectors, with the black curve showing the bulk average. Their accumulation near one boundary demonstrates the NHSE of the bulk rapidity modes. The figure therefore separates the spectral edge rapidities from the skin-localized bulk modes. Parameters are \(t=1\), \(d=0.5\), \(\mu=0\), \(\gamma_\ell=\gamma_g=0.2\), and \(N=150\).
}
\label{fig:nhse_profiles}
\end{figure}
To visualize the NHSE directly, we compute the right eigenvectors of the finite OBC rapidity matrix \(X_{\rm OBC}\). This is the standard eigenmode diagnostic of the skin effect: in systems with NHSE or Liouvillian skin effect, a macroscopic set of right eigenmodes accumulates near an open boundary \cite{PhysRevLett.123.170401,PhysRevB.111.L140303}. For a normalized right eigenvector \(u_n^R\), we define the unit-cell-resolved weight
\begin{equation}
P_n(j)=
\frac{
\sum_{\alpha=1}^{4}\left|u_n^R(j,\alpha)\right|^2
}{
\sum_{j,\alpha}\left|u_n^R(j,\alpha)\right|^2
},
\label{eq:cell_resolved_weight}
\end{equation}
where \(j\) labels the unit cell and \(\alpha=1,\ldots,4\) labels the four Majorana components
\((w^1_{j,A},w^1_{j,B},w^2_{j,A},w^2_{j,B})\) inside that unit cell. Thus \(P_n(j)\) gives the total right-eigenvector weight of mode \(n\) in unit cell \(j\), summed over both sublattices and both Majorana flavors, with \(\sum_j P_n(j)=1\).

Figure~\ref{fig:nhse_profiles} separates the skin localization of bulk modes from the isolated edge rapidities. The bulk right eigenvectors in Fig.~\ref{fig:nhse_profiles}(b) accumulate near one boundary, demonstrating the NHSE of the rapidity matrix. The modes in Fig.~\ref{fig:nhse_profiles}(a) are the isolated edge rapidities selected from the OBC spectrum. Since NHSE also confines the bulk modes, localization alone does not identify a topological edge rapidity. 

\subsection{Bulk gap closing and exceptional windows}
\label{subsec:exceptional_windows}

We now explain the shaded real-gapless windows in Fig.~\ref{fig:pbc_obc_rapidity_spectra}. Finite bond dissipation does not shift the Hermitian critical centers in Eq.~\eqref{eq:hermitian_critical_points}. Instead, each Hermitian gap closing broadens into a finite interval in which the PBC rapidity spectrum is real-gapless.

\begin{figure}[t]
    \centering
    \includegraphics[width=0.87\columnwidth]{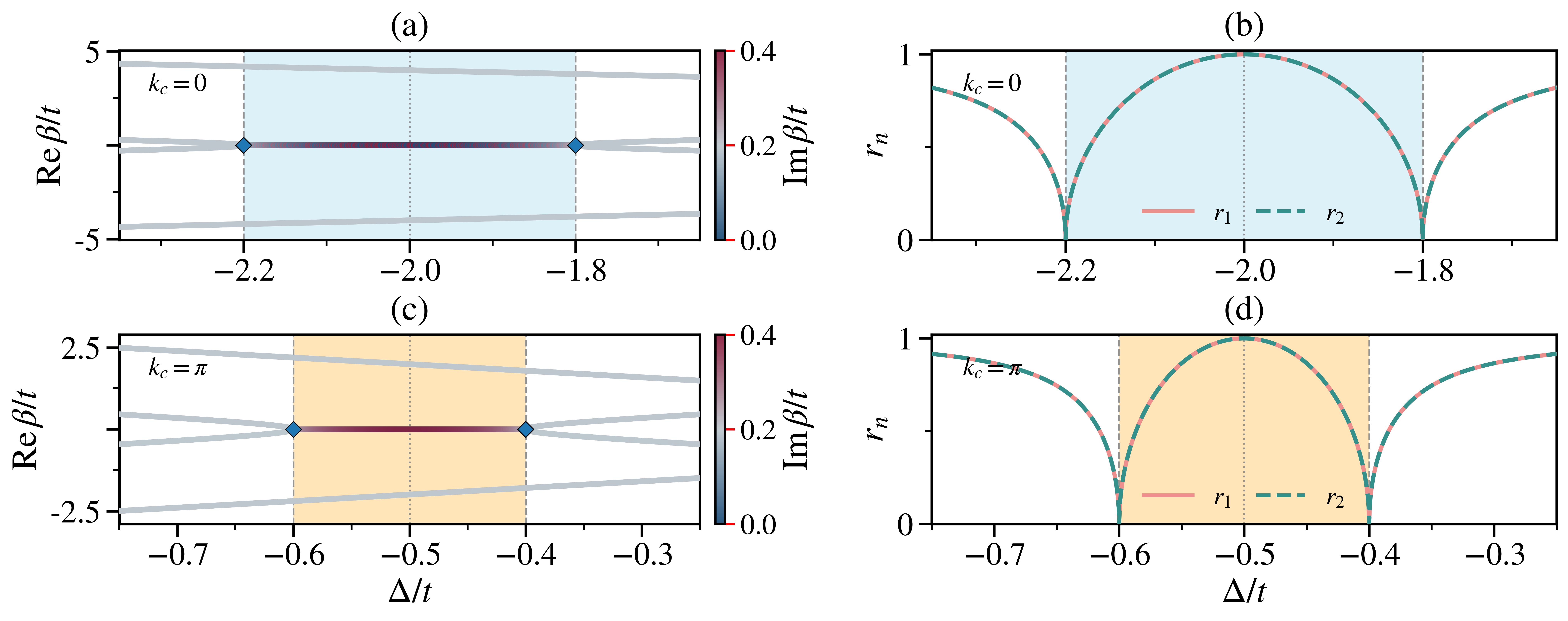}
    \caption{
    Exceptional-point diagnostics for the PBC bulk rapidity spectrum. Parameters are \(t=1\), \(d=0.5\), \(\mu=0\), and balanced gain and loss \(\gamma_\ell=\gamma_g=0.2\), so that \(\Gamma=0.4\). (a) Real parts of the four rapidities obtained by diagonalizing the full \(4\times4\) Bloch rapidity matrix \(X(k)\) at the outer critical momentum \(k_c=0\), plotted as a function of \(\Delta/t\). The color scale denotes \(\mathrm{Im}\,\beta/t\). The shaded region marks the real-gapless window centered at \(\Delta/t=-2\). The diamond markers denote the two analytic boundaries, \(\Delta/t=-2.2\) and \(-1.8\). (b) Biorthogonal phase rigidity \(r_n\) of the two coalescing rapidity modes at \(k_c=0\). The collapse of \(r_n\) at the two window boundaries confirms eigenvector coalescence. (c) Same as (a), but for the inner critical momentum \(k_c=\pi\), with a window centered at \(\Delta/t=-0.5\) and boundaries \(-0.6\) and \(-0.4\). (d) Corresponding phase rigidity for the two coalescing modes at \(k_c=\pi\).
    }
\label{fig:ep_diagnostics}
\end{figure}

The boundaries of these intervals follow from the sector rapidities in Eq.~\eqref{eq:beta_pbc}. Since the shift \(i\Gamma/2\) is purely imaginary, \(\mathrm{Re}\,\beta=0\) is controlled by the square-root term. Thus a rapidity lies on the line \(\mathrm{Re}\,\beta=0\) when \(R_\eta(e^{ik})\) is real and non-positive. On the Bloch unit circle,
\begin{equation}
R_\eta(e^{ik})
=
a_\eta^2+b_\eta^2+2a_\eta b_\eta\cos k-\frac{\Gamma^2}{4}
+i\Gamma b_\eta\sin k .
\label{eq:Reta_unit_circle}
\end{equation}
For generic \(b_\eta\neq0\), the imaginary part vanishes at the high-symmetry momenta \(k_c=0,\pi\).

At \(k_c=0\),
\begin{equation}
R_\eta(1)=\left(a_\eta+b_\eta\right)^2-\frac{\Gamma^2}{4}.
\label{eq:outer_R}
\end{equation}
The rapidity spectrum is real-gapless when
\begin{equation}
|a_\eta+b_\eta|\leq\frac{\Gamma}{2}.
\label{eq:outer_condition}
\end{equation}
Using \(a_\eta+b_\eta=2(t+\eta d\Delta)\), the outer-window boundaries are
\begin{equation}
\frac{\Delta}{t}
=
-\frac{\eta}{d}
\pm
\frac{\Gamma}{4td},
\qquad
k_c=0 .
\label{eq:outer_window_boundaries}
\end{equation}
At \(k_c=\pi\),
\begin{equation}
R_\eta(-1)=\left(a_\eta-b_\eta\right)^2-\frac{\Gamma^2}{4}.
\label{eq:inner_R}
\end{equation}
The rapidity spectrum is real-gapless when
\begin{equation}
|a_\eta-b_\eta|\leq\frac{\Gamma}{2}.
\label{eq:inner_condition}
\end{equation}
Using \(a_\eta-b_\eta=2(td-\eta\Delta)\), the inner-window boundaries are
\begin{equation}
\frac{\Delta}{t}
=
\eta d
\pm
\frac{\Gamma}{4t},
\qquad
k_c=\pi .
\label{eq:inner_window_boundaries}
\end{equation}
For \(t=1\), \(d=0.5\), and \(\Gamma=0.4\), these formulas give the outer windows \(-2.2<\Delta/t<-1.8\) and \(1.8<\Delta/t<2.2\), and the inner windows \(-0.6<\Delta/t<-0.4\) and \(0.4<\Delta/t<0.6\). We verify that the boundaries of these real-gapless windows are exceptional points of the PBC rapidity matrix. The numerical data in Fig.~\ref{fig:ep_diagnostics} are obtained from the full \(4\times4\) Bloch rapidity matrix \(X(k)\), not from a reduced sector matrix. The sector expressions above are used only to obtain the analytic boundary positions.

The phase rigidity is defined as \cite{PhysRevA.93.042116}
\begin{equation}
r_n=
\frac{
|\langle u_n^L|u_n^R\rangle|
}{
\sqrt{
\langle u_n^L|u_n^L\rangle
\langle u_n^R|u_n^R\rangle
}
},
\label{eq:phase_rigidity}
\end{equation}
where
\begin{equation}
X(k_c)u_n^R=\beta_nu_n^R,
\qquad
X^\dagger(k_c)u_n^L=\beta_n^*u_n^L .
\end{equation}
For an ordinary nondegenerate mode, \(r_n\) is finite. At an exceptional point, the left and right eigenvectors become self-orthogonal and \(r_n\to0\). Figures~\ref{fig:ep_diagnostics}(b) and \ref{fig:ep_diagnostics}(d) show that the phase rigidity of the two participating modes collapses at the analytic window boundaries. Together with the rapidity coalescence in Figs.~\ref{fig:ep_diagnostics}(a) and \ref{fig:ep_diagnostics}(c), this confirms that the real-gapless windows are bounded by second-order exceptional points.

At a window boundary, the participating sector block becomes defective explicitly. At the critical momentum, the shifted sector block has the form
\begin{equation}
    X_\eta(e^{ik_c})-\frac{i\Gamma}{2}\mathbf{1}_2
    =
    \begin{pmatrix}
        0 & u_c\\
        0 & 0
    \end{pmatrix},
    \qquad u_c\neq0,
\end{equation}
or the transposed form with a single nonzero lower off-diagonal element. Hence the degenerate eigenvalue has algebraic multiplicity two but only one independent eigenvector, confirming its exceptional-point character.

\section{Sector non-Bloch topology and edge rapidities}
\label{sec:sector_nonbloch}
\label{subsec:sector_gbz}
\begin{figure}[t]
    \centering
    \includegraphics[width=0.65\textwidth]{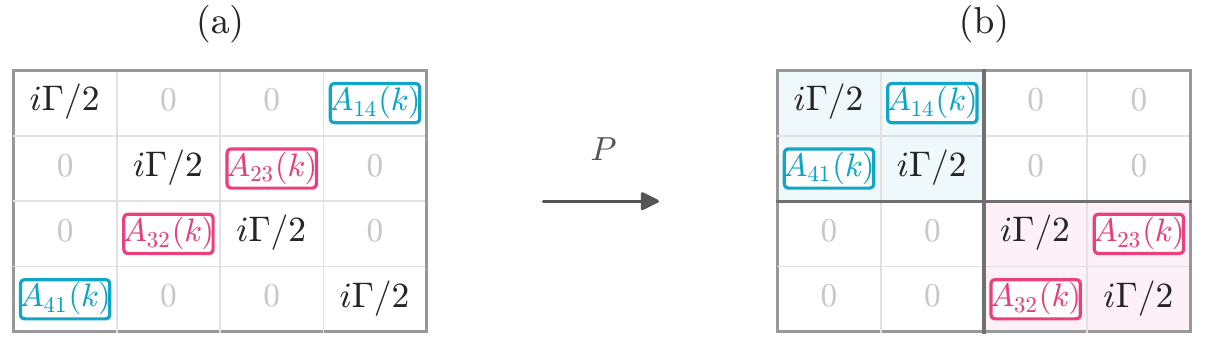}
    \caption{Matrix-level schematic of the sector decomposition at
    \(\mu=0\). In the original ordering
    \((w^1_A,w^1_B,w^2_A,w^2_B)\), the only off-diagonal couplings are
    between the pairs \((1,4)\) and \((2,3)\). The permutation
    \((1,2,3,4)\to(1,4,2,3)\) makes the direct-sum structure
    \(X_{14}(k)\oplus X_{23}(k)\) explicit.}
    \label{figS:matrix_reordering}
\end{figure}

The spectral comparison above shows that the OBC bulk rapidity spectrum is not obtained from the ordinary Bloch contour. We now derive the sector non-Bloch construction that explains the OBC continua and the isolated edge rapidities. At \(\mu=0\), the Bloch rapidity matrix in Eq.~\eqref{eq:Xk} separates into two independent Majorana sectors. In the local ordering
\((w^1_A,w^1_B,w^2_A,w^2_B)\), it takes the form
\begin{equation}
X(k)=
\begin{pmatrix}
 i\Gamma/2 & 0 & 0 & A_{14}(k)\\
 0 & i\Gamma/2 & A_{23}(k) & 0\\
 0 & A_{32}(k) & i\Gamma/2 & 0\\
 A_{41}(k) & 0 & 0 & i\Gamma/2
\end{pmatrix},
\label{eq:app_X_mu0_matrix}
\end{equation}
where the nonzero off-diagonal elements are
\[
\begin{aligned}
A_{14}(k)
&=
i\left(t_1-\Delta_1+\frac{\Gamma}{2}\right)
+
i\left(t_2+\Delta_2\right)e^{-ik},\\
A_{41}(k)
&=
i\left(\Delta_1-t_1+\frac{\Gamma}{2}\right)
-
i\left(t_2+\Delta_2\right)e^{ik},\\
A_{23}(k)
&=
i\left(t_1+\Delta_1-\frac{\Gamma}{2}\right)
+
i\left(t_2-\Delta_2\right)e^{ik},\\
A_{32}(k)
&=
-i\left(t_1+\Delta_1+\frac{\Gamma}{2}\right)
-
i\left(t_2-\Delta_2\right)e^{-ik}.
\end{aligned}
\]
The corresponding permutation to the sector basis is shown in
Fig.~\ref{figS:matrix_reordering}.
Thus the two sectors are
\[
    \eta=+:\;(w^1_A,w^2_B),
    \qquad
    \eta=-:\;(w^1_B,w^2_A).
\]
Equivalently, these are the \(14\) and \(23\) sectors in the local
Majorana ordering.

We define
\begin{equation}
    a_\eta=t_1-\eta\Delta_1,
    \qquad
    b_\eta=t_2+\eta\Delta_2,
    \qquad
    \eta=\pm .
\label{eq:app_aeta_beta}
\end{equation}
With \(z=e^{ik}\) under PBC, both sectors can be written as
\begin{equation}
    X_\eta(z)
    =
    \frac{i\Gamma}{2}\mathbf{I}_2
    +
    \begin{pmatrix}
        0 & u_\eta(z)\\
        v_\eta(z) & 0
    \end{pmatrix},
\label{eq:app_Xeta_z}
\end{equation}
where
\[
\begin{aligned}
u_\eta(z)
&=
i\Bigl(
a_\eta+\eta\frac{\Gamma}{2}
+b_\eta z^{-\eta}
\Bigr),
\\
v_\eta(z)
&=
-i\Bigl(
a_\eta-\eta\frac{\Gamma}{2}
+b_\eta z^{\eta}
\Bigr).
\end{aligned}
\]
The two rapidities in sector \(\eta\) are
\begin{equation}
    \beta_{\eta,\pm}^{\rm PBC}(k)
    =
    \frac{i\Gamma}{2}
    \pm
    \sqrt{R_\eta(e^{ik})},
\label{eq:app_beta_eta_pbc}
\end{equation}
where
\begin{equation}
    R_\eta(z)
    =
    \left(
    a_\eta+\frac{\Gamma}{2}+b_\eta z^{-1}
    \right)
    \left(
    a_\eta-\frac{\Gamma}{2}+b_\eta z
    \right).
\label{eq:app_Reta_z}
\end{equation}
On the Bloch unit circle,
\begin{equation}
    R_\eta(e^{ik})
    =
    a_\eta^2+b_\eta^2
    +2a_\eta b_\eta\cos k
    -\frac{\Gamma^2}{4}
    +i\Gamma b_\eta\sin k .
\label{eq:app_Reta_k}
\end{equation}

We now obtain the sector-dependent GBZ. For a fixed rapidity
\(\beta\), define
\[
    q_\eta=
    \left(
    \beta-\frac{i\Gamma}{2}
    \right)^2 .
\]
The characteristic equation \(q_\eta=R_\eta(z)\) becomes, after
multiplication by \(z\),
\[
b_\eta\left(a_\eta+\frac{\Gamma}{2}\right)z^2
+
\left[
a_\eta^2+b_\eta^2-\frac{\Gamma^2}{4}
-q_\eta
\right]z
+
b_\eta\left(a_\eta-\frac{\Gamma}{2}\right)
=0 .
\]
Let \(z_1\) and \(z_2\) be the two roots. For a nearest-neighbor
non-Hermitian two-band chain, the GBZ condition is
\(|z_1|=|z_2|\) \cite{PhysRevLett.121.086803,PhysRevLett.123.066404,PhysRevLett.124.086801}. Since
\[
    z_1z_2
    =
    \frac{
    a_\eta-\Gamma/2
    }{
    a_\eta+\Gamma/2
    },
\]
the GBZ radius of sector \(\eta\) is
\begin{equation}
    r_\eta
    =
    \sqrt{
    \left|
    \frac{a_\eta-\Gamma/2}{a_\eta+\Gamma/2}
    \right|
    } .
\label{eq:app_gbz_radius}
\end{equation}
The non-Bloch substitution is \(z=r_\eta e^{ik}\), with
\(k\in[0,2\pi)\). The OBC bulk rapidities are therefore
\begin{equation}
    \beta_{\eta,\pm}^{\rm GBZ}(k)
    =
    \frac{i\Gamma}{2}
    \pm
    \sqrt{
    R_\eta(r_\eta e^{ik})
    } .
\label{eq:app_beta_gbz}
\end{equation}
The full OBC bulk continuum is the union of the two sector continua,
\[
    \mathrm{Spec}_{\rm OBC}^{\rm bulk}
    =
    \bigcup_{\eta=\pm}
    \left\{
    \beta_{\eta,\pm}^{\rm GBZ}(k):
    k\in[0,2\pi)
    \right\}.
\]
Thus the dissipative dimerized Kitaev chain has two independent
Majorana sectors and, in general, two distinct GBZ radii.

When \(a_\eta^2>\Gamma^2/4\), Eq.~\eqref{eq:app_beta_gbz} can be
written in a more transparent form. Defining
\[
    a_{\eta,{\rm eff}}
    =
    \sqrt{
    a_\eta^2-\frac{\Gamma^2}{4}
    },
\]
one obtains
\[
    R_\eta(r_\eta e^{ik})
    =
    a_{\eta,{\rm eff}}^2
    +
    b_\eta^2
    +
    2a_{\eta,{\rm eff}}b_\eta\cos k ,
\]
up to the sign fixed by the branch choice of the square root. Hence
\[
    \beta_{\eta,\pm}^{\rm GBZ}(k)
    =
    \frac{i\Gamma}{2}
    \pm
    \sqrt{
    a_{\eta,{\rm eff}}^2
    +
    b_\eta^2
    +
    2a_{\eta,{\rm eff}}b_\eta\cos k
    } .
\]
This form shows explicitly that the OBC bulk bands are shifted by
\(i\Gamma/2\), while the non-Bloch continuum is governed by an
effective intracell coupling \(a_{\eta,{\rm eff}}\).

\subsection{Non-Bloch winding numbers}
\label{app:sector_winding_derivation}

We now derive the sector winding criterion used in the main text. The sector matrices and their GBZs have already been obtained above. From Eq.~\eqref{eq:app_Xeta_z}, we write
\begin{equation}
    u_\eta(z)=i h_{\eta,+}(z),
    \qquad
    v_\eta(z)=-i h_{\eta,-}(z),
\end{equation}
where
\begin{align}
    h_{\eta,+}(z)
    &=
    a_\eta+\eta\frac{\Gamma}{2}
    +b_\eta z^{-\eta},
    \nonumber\\
    h_{\eta,-}(z)
    &=
    a_\eta-\eta\frac{\Gamma}{2}
    +b_\eta z^{\eta}.
\label{eq:app_h_eta_pm}
\end{align}
Here \(\eta=+\) denotes the \(1\!-\!4=(w_A^1,w_B^2)\) sector and \(\eta=-\) denotes the \(2\!-\!3=(w_B^1,w_A^2)\) sector.

The uniform shift \(i\Gamma/2\) does not affect the winding. We therefore introduce
\begin{equation}
    \widetilde X_\eta(z)
    =
    X_\eta(z)-\frac{i\Gamma}{2}\mathbf{1}_2
    =
    \begin{pmatrix}
        0 & i h_{\eta,+}(z)\\
        -i h_{\eta,-}(z) & 0
    \end{pmatrix}.
\label{eq:app_Xeta_shifted}
\end{equation}
This matrix has the same off-diagonal form as a non-Hermitian SSH chain. The sector invariant is therefore the non-Bloch chiral winding of the two off-diagonal functions, evaluated on the sector GBZ rather than on the ordinary Brillouin zone~\cite{PhysRevLett.121.086803,PhysRevLett.123.066404,PhysRevLett.124.086801}. Following the standard non-Bloch SSH construction, we define
\begin{equation}
    w_{\eta,\pm}
    =
    \frac{1}{2\pi i}
    \oint_{\mathcal C_\eta}
    d\log h_{\eta,\pm}(z),
\label{eq:app_w_eta_pm}
\end{equation}
and the sector winding as the relative winding \cite{PhysRevLett.123.066404}
\begin{equation}
    W_\eta
    =
    \frac{1}{2}
    \left(
    w_{\eta,-}-w_{\eta,+}
    \right)
    =
    \frac{1}{4\pi i}
    \oint_{\mathcal C_\eta}
    d\log
    \frac{h_{\eta,-}(z)}{h_{\eta,+}(z)} .
\label{eq:app_W_eta_def}
\end{equation}
This is the sector-resolved analogue of the non-Bloch SSH winding. The integer plotted in the main text is the nonnegative winding
\begin{equation}
    \nu_\eta=|W_\eta|.
\end{equation}

The explicit criterion for \(\nu_\eta\) follows by evaluating Eq.~\eqref{eq:app_W_eta_def} with the argument principle. We first consider \(\eta=+\). In this sector,
\begin{align}
    h_{+,+}(z)
    &=
    a_+ + \frac{\Gamma}{2}
    +
    b_+ z^{-1},
    \nonumber\\
    h_{+,-}(z)
    &=
    a_+ - \frac{\Gamma}{2}
    +
    b_+ z .
\end{align}
The function \(h_{+,+}\) has one pole at \(z=0\) and one zero at
\begin{equation}
    z_{+,+}^{(0)}
    =
    -\frac{b_+}{a_+ + \Gamma/2},
\end{equation}
whereas \(h_{+,-}\) has no pole and one zero at
\begin{equation}
    z_{+,-}^{(0)}
    =
    -\frac{a_+ - \Gamma/2}{b_+}.
\end{equation}
The sector GBZ is the circle \(|z|=r_+\), with
\begin{equation}
    r_+
    =
    \sqrt{
    \left|
    \frac{a_+-\Gamma/2}{a_++\Gamma/2}
    \right|
    } .
\end{equation}
The zero of \(h_{+,-}\) lies inside the GBZ when
\begin{equation}
    \left|
    \frac{a_+-\Gamma/2}{b_+}
    \right|
    < r_+ ,
\end{equation}
which is equivalent to
\begin{equation}
    |b_+|^2
    >
    \left|
    a_+^2-\frac{\Gamma^2}{4}
    \right| .
\label{eq:app_plus_top_condition}
\end{equation}
In the same parameter regime, the zero of \(h_{+,+}\) lies outside the GBZ, while its pole at \(z=0\) remains inside. Hence the relative winding is nonzero and \(|W_+|=1\). In the opposite regime,
\begin{equation}
    |b_+|^2
    <
    \left|
    a_+^2-\frac{\Gamma^2}{4}
    \right| ,
\end{equation}
the zero and pole contributions cancel, giving \(|W_+|=0\).

The \(\eta=-\) sector is obtained by interchanging \(z\) and \(z^{-1}\), which reverses the orientation of the winding but not its absolute value. Therefore both sectors obey
\begin{equation}
    \nu_\eta
    =
    \begin{cases}
    1, &
    |b_\eta|^2>
    \left|a_\eta^2-\Gamma^2/4\right|,
    \\[3pt]
    0, &
    |b_\eta|^2<
    \left|a_\eta^2-\Gamma^2/4\right|.
    \end{cases}
\label{eq:app_sector_winding_criterion}
\end{equation}
At equality, a zero crosses the sector GBZ and the shifted point gap closes at \(\beta=i\Gamma/2\). The winding number is therefore not assigned at equality.

At \(\mu=0\), the finite OBC rapidity matrix is the direct sum of the two sector OBC matrices after the same Majorana-sector reordering. By the non-Bloch bulk-boundary correspondence applied to each sector chain, a nonzero \(\nu_\eta\) gives one isolated edge rapidity at each end of the open chain. Therefore
\begin{equation}
    n_{\rm edge}^{\rm pred}
    =
    2(\nu_{14}+\nu_{23}).
\label{eq:app_edge_count}
\end{equation}
The factor of two counts the two physical boundaries. This is the edge-count prediction used in the main text.

\section{Trivial control and entanglement-spectrum zero-event scan}
\label{secS:ES_scan}
\begin{figure}[t]
    \centering
    \includegraphics[width=0.95\columnwidth]{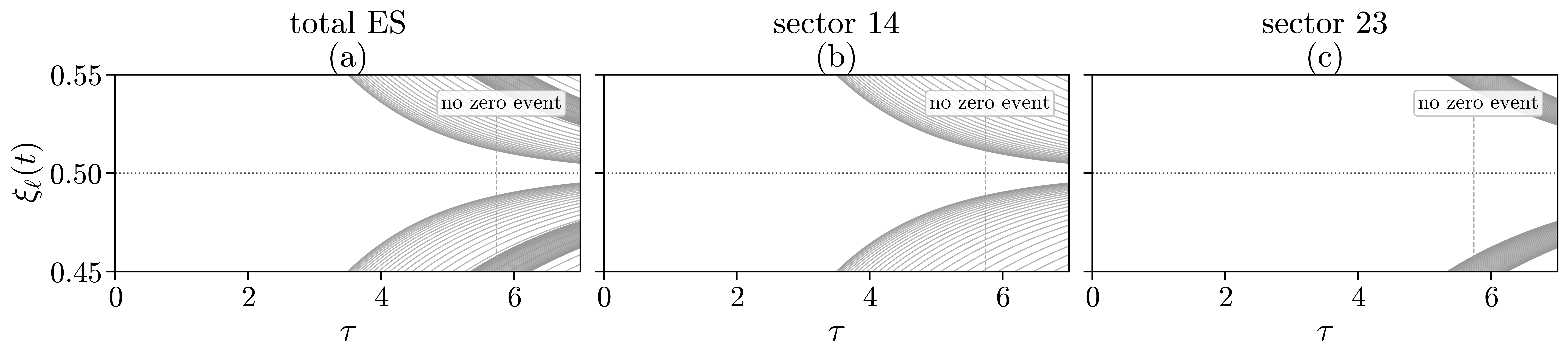}
    \caption{
    Trivial-to-trivial control quench for the entanglement-spectrum dynamics under PBC. The initial state is the trivial Hermitian ground state at \(\Delta_i/t=0.25\), and the post-quench value is also \(\Delta/t=0.25\), with balanced gain and loss turned on at \(\tau=0\). Since \((\nu_{14},\nu_{23})=(0,0)\), the total ES, sector-$1\!-\!4$ covariance ES, and sector-$2\!-\!3$ covariance ES show no finite-time zero event at \(\xi_\ell=1/2\). Parameters are \(N=150\), \(N_S=75\), \(t=1\), \(d=0.5\), \(\mu=0\), and \(\gamma_\ell=\gamma_g=0.2\).
    }
    \label{figS:ES_trivial_control}
\end{figure}
\begin{figure}[t]
    \centering
    \includegraphics[width=0.8\columnwidth]{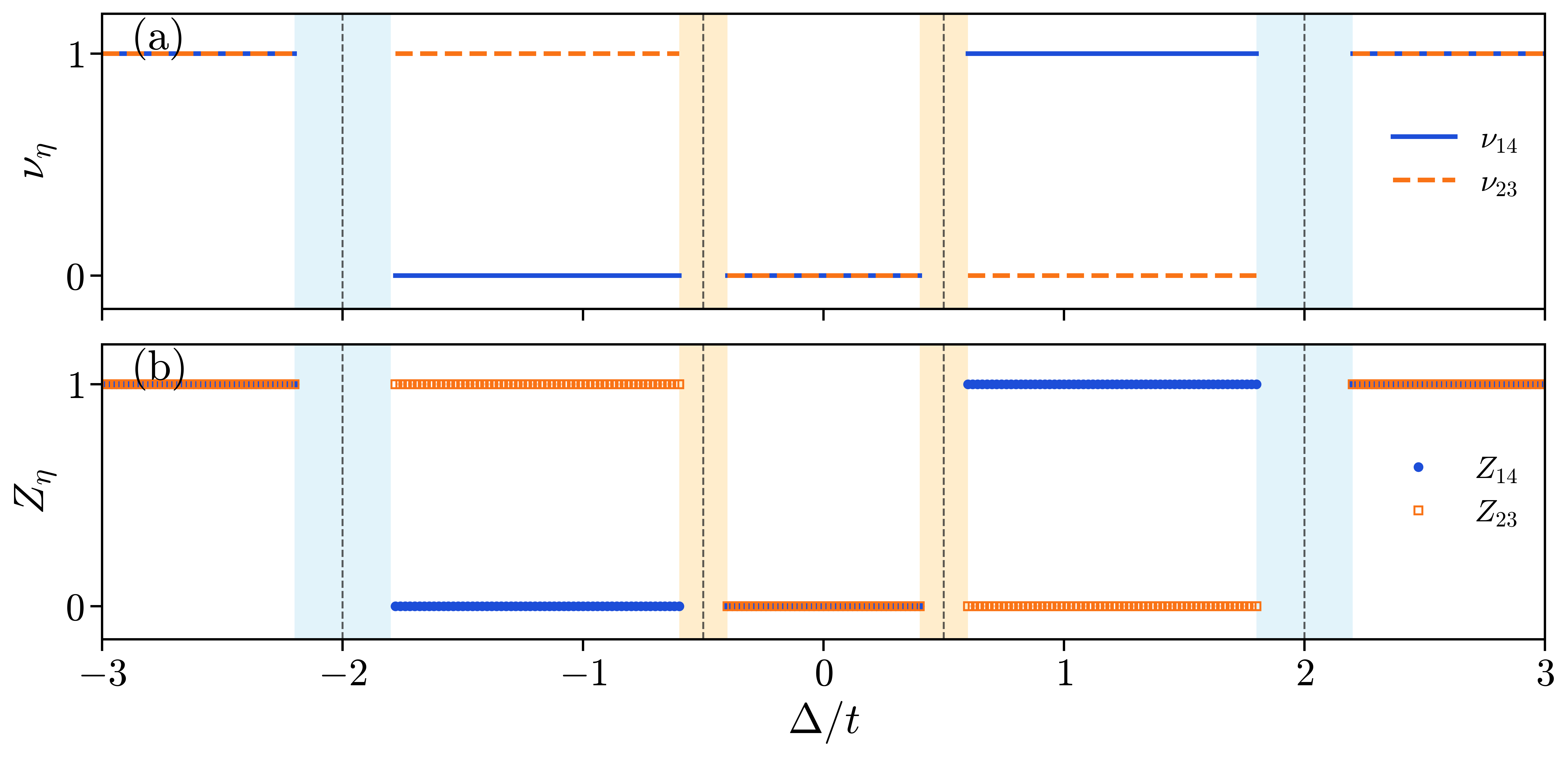}
    \caption{
    Sector-resolved ES zero-event scan. (a) Sector non-Bloch winding numbers \(\nu_{14}\) and \(\nu_{23}\) as functions of the post-quench pairing \(\Delta/t\). (b) Binary ES zero-event indicators \(Z_{14}\) and \(Z_{23}\), defined in Eq.~\eqref{eqS:ZES_def}, extracted from PBC Lindblad evolution. The shaded regions denote real-gapless rapidity windows, where the finite-time ES-winding correspondence is not used as a sharp diagnostic. Away from these windows and transition points, \(Z_\eta=\nu_\eta\), showing that ES zero events resolve the sector whose post-quench non-Bloch winding is nonzero. Parameters are \(N=150\), \(N_S=75\), \(t=1\), \(d=0.5\), \(\mu=0\), and \(\gamma_\ell=\gamma_g=0.2\).
    }
    \label{figS:ES_zero_event_scan}
\end{figure}
The main text shows representative entanglement-spectrum dynamics for quenches into nontrivial regimes. Here we perform two checks. We first show a trivial-to-trivial control quench, where no finite-time entanglement zero event occurs. We then scan the post-quench pairing strength and compare the sector-resolved zero-event indicators with the sector non-Bloch winding numbers. The initial state is fixed throughout this section. It is the ground state of the closed Hermitian chain with
\[
t=1,\qquad d=0.5,\qquad \mu=0,\qquad \Delta_i/t=0.25,
\qquad \gamma_\ell=\gamma_g=0 .
\]
At \(\tau=0^+\), the pairing is changed to the post-quench value \(\Delta\), and balanced bond dissipation with \(\gamma_\ell=\gamma_g=0.2\) is turned on. The Lindblad evolution is performed under PBC. We use a subsystem of \(N_S=75\) consecutive unit cells in a chain of \(N=150\) unit cells. Throughout this section, \(\Delta/t\) denotes the post-quench value.

For each evolution, we compute the total subsystem ES and the sector-restricted covariance ES diagnostics. The total ES is obtained by keeping all four Majorana components in the subsystem. The sector-$1\!-\!4$ diagnostic is obtained by keeping only \((w_A^1,w_B^2)\), while the sector-$2\!-\!3$ diagnostic is obtained by keeping only \((w_B^1,w_A^2)\). These sector-restricted spectra are not separate physical bipartitions; they are diagnostics used to identify which Majorana sector produces a zero event. Figure~\ref{figS:ES_trivial_control} shows the trivial-to-trivial control quench. The post-quench value is \(\Delta/t=0.25\), for which the sector windings are \((\nu_{14},\nu_{23})=(0,0)\). The total ES and both sector-restricted ES diagnostics show no finite-time zero event at \(\xi_\ell=1/2\). This confirms that the zero events in the nontrivial quenches are not artifacts of the final balanced-dissipation approach to \(\xi_\ell=1/2\), but are tied to nonzero post-quench sector winding.

We now scan the post-quench pairing \(\Delta/t\). For each value of \(\Delta/t\), we define a binary zero-event indicator
\begin{equation}
Z_\eta(\Delta)=
\begin{cases}
1,& \text{if the sector-\(\eta\) covariance ES has a finite-time zero event at } \xi_\ell=1/2,\\
0,& \text{otherwise},
\end{cases}
\qquad \eta=14,23 .
\label{eqS:ZES_def}
\end{equation}
Only finite-time zero events are counted. The final asymptotic approach to \(\xi_\ell=1/2\), which occurs because balanced gain and loss drives the covariance matrix to a featureless steady state, is not counted as a topological event. Figure~\ref{figS:ES_zero_event_scan} compares \(Z_{14}\) and \(Z_{23}\) with the sector non-Bloch winding numbers \(\nu_{14}\) and \(\nu_{23}\). Away from transition points and the shaded real-gapless rapidity windows, the indicators follow the sector windings:
\begin{equation}
Z_{14}(\Delta)=\nu_{14}(\Delta),
\qquad
Z_{23}(\Delta)=\nu_{23}(\Delta).
\label{eqS:ZES_rule}
\end{equation}
Since the initial state is trivial in both sectors, this shows that a finite-time ES zero event occurs precisely in the sector whose post-quench non-Bloch winding is nonzero.

\section{Finite chemical potential: full-matrix spectra and total ES}
\label{secS:finite_mu}
\begin{figure*}[t]
    \centering
    \includegraphics[width=0.98\textwidth]{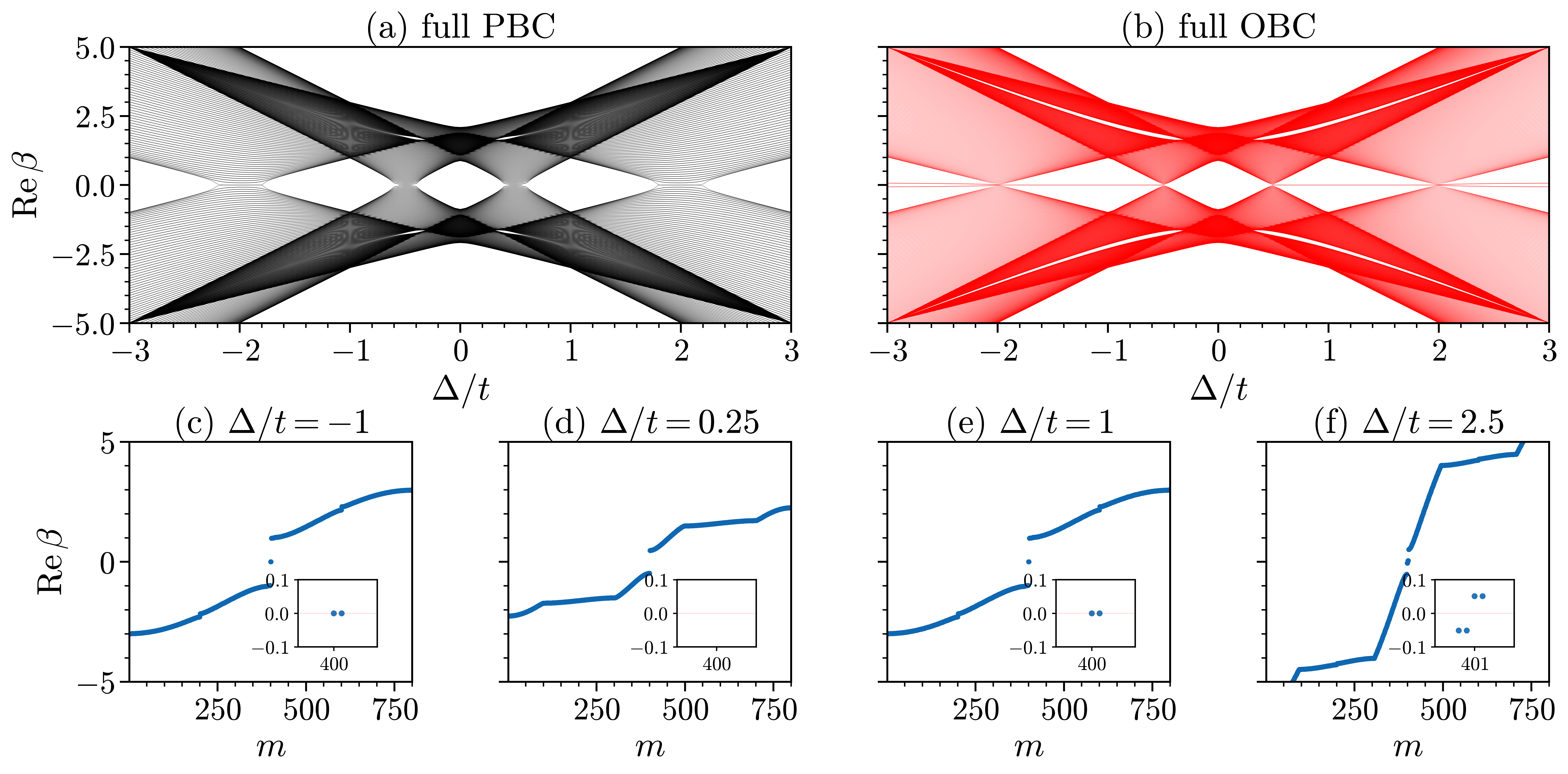}
    \caption{
    Full rapidity spectra at finite chemical potential, \(\mu/t=0.1\). 
    (a) PBC spectrum obtained by diagonalizing the full Bloch rapidity matrix \(X(k)\). 
    (b) OBC spectrum obtained by diagonalizing the finite-chain matrix \(X_{\rm OBC}\). 
    In panels (a) and (b), we plot \({\rm Re}\,\beta\) as a function of \(\Delta/t\). 
    (c)--(f) Fixed-\(\Delta/t\) OBC spectra at \(\Delta/t=-1,0.25,1,\) and \(2.5\), respectively. The real parts of the OBC rapidities are sorted and plotted versus the mode index \(m\), and the insets zoom near \({\rm Re}\,\beta=0\). 
    Since finite \(\mu\) hybridizes the $1\!-\!4$ and $2\!-\!3$ Majorana sectors, no sector GBZ or sector winding is assigned.
    Parameters are \(t=1\), \(d=0.5\), \(\gamma_\ell=\gamma_g=0.2\), and \(\mu/t=0.1\).
    }
    \label{figS:finite_mu_rebeta}
\end{figure*}
The sector-resolved construction used in the main text relies on the exact Majorana block structure at \(\mu=0\). At this point, the rapidity matrix separates into the \(1\!-\!4=(w_A^1,w_B^2)\) and \(2\!-\!3=(w_B^1,w_A^2)\) sectors, each with its own GBZ and non-Bloch winding. A finite chemical potential adds the term \(-\mu\tau_y\sigma_0\) to the Bloch rapidity matrix and couples these two sectors. Consequently, for \(\mu\neq0\), the sector GBZs and sector windings are no longer independent topological diagnostics. We therefore treat finite \(\mu\) only as a full-matrix crossover check, using the full Bloch rapidity matrix \(X(k)\), the finite-chain matrix \(X_{\rm OBC}\), and the total ES of the spatial subsystem.

Figure~\ref{figS:finite_mu_rebeta} shows the full rapidity spectra at \(\mu/t=0.1\). Panels (a) and (b) plot \({\rm Re}\,\beta\) as a function of \(\Delta/t\), obtained by direct diagonalization of \(X(k)\) under PBC and \(X_{\rm OBC}\) under OBC, respectively. No sector GBZ curve is used. The PBC and OBC spectra remain visibly different, showing that the rapidity spectrum remains boundary-condition sensitive after sector mixing. Panels (c)--(f) show fixed-\(\Delta/t\) OBC spectra, where the real parts of the OBC rapidities are sorted and plotted versus the mode index \(m\). The insets zoom near the central rapidity gap, \({\rm Re}\,\beta=0\). At the trivial point \(\Delta/t=0.25\), no isolated central rapidity appears. At \(\Delta/t=\pm1\), a pair of central in-gap rapidities remains visible. In the SSH-like regime, \(\Delta/t=2.5\), the central rapidities that are degenerate at \(\mu=0\) split away from \({\rm Re}\,\beta=0\). This splitting is consistent with hybridization of the two sector edge modes by finite \(\mu\) \cite{Zhang2025StaggeredKitaev}. Since the sectors are no longer independent, these finite-\(\mu\) data are not assigned sector windings or sector edge counts.

\begin{figure*}[t]
    \centering
    \includegraphics[width=0.98\textwidth]{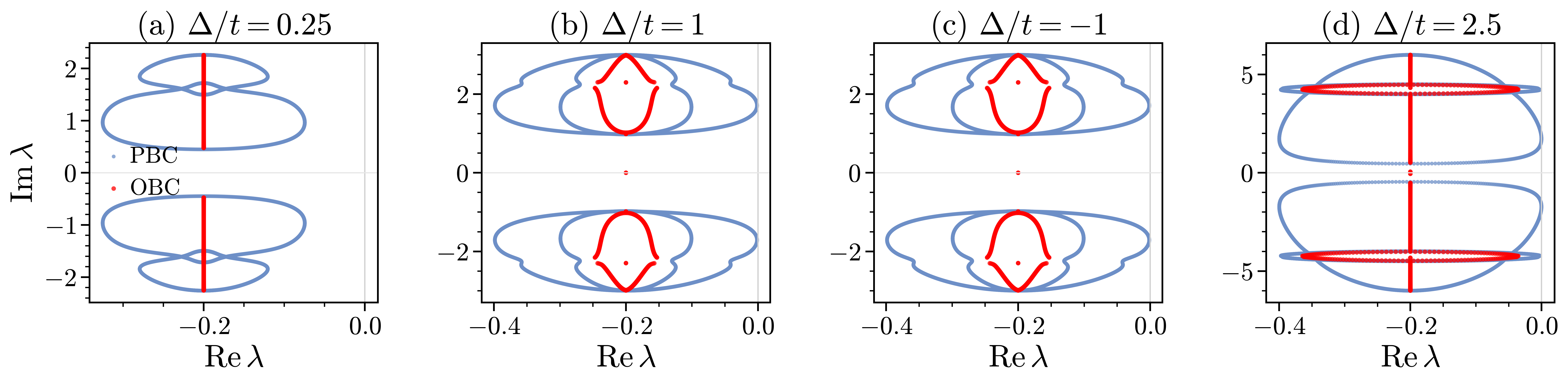}
    \caption{
    Full PBC and OBC spectra in the damping plane \(\lambda=i\beta\) at finite chemical potential, \(\mu/t=0.1\). The four panels correspond to \(\Delta/t=0.25,1,-1,\) and \(2.5\). Blue points denote the full PBC spectrum obtained from \(X(k)\), and red points denote the full OBC spectrum obtained from \(X_{\rm OBC}\). The distance of the blue spectrum from the line \({\rm Re}\,\lambda=0\) gives the full PBC Liouvillian gap. The trivial point \(\Delta/t=0.25\) is Liouvillian-gapped, while the spectra for \(\Delta/t=1,-1,\) and \(2.5\) reach \({\rm Re}\,\lambda=0\), indicating a closed full PBC Liouvillian gap. Parameters are \(t=1\), \(d=0.5\), \(\gamma_\ell=\gamma_g=0.2\), and \(\mu/t=0.1\).
    }
    \label{figS:finite_mu_lambda}
\end{figure*}

\begin{figure*}[t]
    \centering
    \includegraphics[width=0.98\textwidth]{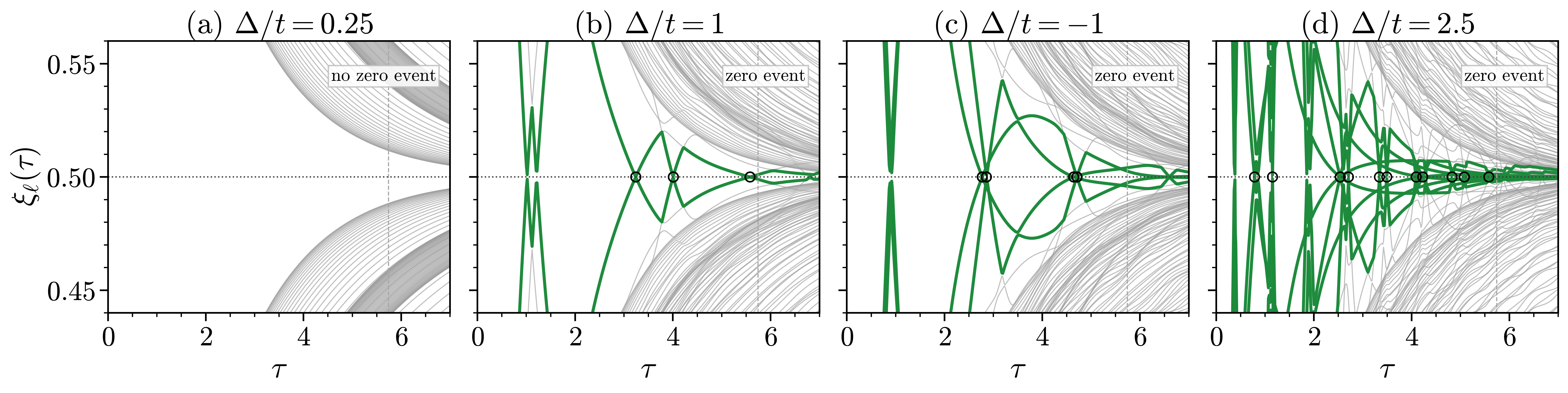}
    \caption{
    Total entanglement-spectrum dynamics at finite chemical potential, \(\mu/t=0.1\). 
    Since finite \(\mu\) hybridizes the $1\!-\!4$ and $2\!-\!3$ Majorana sectors, only the total ES is shown. 
    The plotted quantities are the entanglement occupation eigenvalues \(\xi_\ell(\tau)\), with \(\xi_\ell=1/2\) marking an entanglement zero mode. 
    The panels correspond to post-quench values \(\Delta/t=0.25,1,-1,\) and \(2.5\). 
    The trivial control at \(\Delta/t=0.25\) shows no finite-time zero event, whereas the other three quenches show finite-time zero events. 
    Colored curves highlight branches participating in zero events, gray curves denote the remaining branches, and open circles mark zero events at \(\xi_\ell=1/2\).
    Parameters are \(N=200\), \(N_S=100\), \(t=1\), \(d=0.5\), \(\gamma_\ell=\gamma_g=0.2\), and \(\mu/t=0.1\).
    }
    \label{figS:finite_mu_total_es}
\end{figure*}

The same finite-\(\mu\) spectra are shown in the damping plane \(\lambda=i\beta\) in Fig.~\ref{figS:finite_mu_lambda}. This representation is useful for reading the full PBC Liouvillian gap from the distance of the PBC spectrum to the line \({\rm Re}\,\lambda=0\). With eigenvalues \(\lambda_n(k)\) of \(iX(k)\), we define
\[
    \Lambda_{\rm PBC}^{\rm full}
    =
    \min_{n,k}\left[-2\,{\rm Re}\,\lambda_n(k)\right].
\]
At \(\Delta/t=0.25\), the blue PBC spectrum remains separated from \({\rm Re}\,\lambda=0\), so \(\Lambda_{\rm PBC}^{\rm full}\) is finite. At \(\Delta/t=1,-1,\) and \(2.5\), the blue PBC spectrum reaches \({\rm Re}\,\lambda=0\), showing that the full PBC Liouvillian gap closes. These three values are the finite-\(\mu\) counterparts of the nontrivial \(\mu=0\) cases used in the main text, but no sector winding is assigned at \(\mu/t=0.1\) because the $1\!-\!4$ and $2\!-\!3$ sectors are hybridized. The red OBC spectra are obtained by direct diagonalization of the full \(X_{\rm OBC}\), not from sector GBZs. Thus Fig.~\ref{figS:finite_mu_lambda} shows that the Liouvillian-gap structure and the PBC/OBC spectral distinction persist under weak sector mixing, while the exact sector-resolved interpretation is lost.

We finally test the ES response at finite chemical potential. Since \(\mu/t=0.1\) mixes the two Majorana sectors, the sector-selection rule of the main text is no longer exact. We therefore compute only the total ES, using all Majoranas in the spatial subsystem. The initial state is the ground state of the isolated Hermitian chain with
\[
(t,d,\Delta_i/t,\mu/t,\gamma_\ell,\gamma_g)=(1,0.5,0.25,0.1,0,0).
\]

At \(\tau=0^+\), the pairing is changed to the post-quench value \(\Delta\), and balanced bond dissipation with \(\gamma_\ell=\gamma_g=0.2\) is turned on. The Lindblad evolution is performed under PBC. Throughout this subsection, \(\Delta/t\) denotes the post-quench value.

Figure~\ref{figS:finite_mu_total_es} shows the total ES for four representative post-quench values. The trivial control, \(\Delta/t=0.25\), shows no finite-time zero event. By contrast, the quenches to \(\Delta/t=1\), \(-1\), and \(2.5\), which are continuously connected to the nontrivial \(\mu=0\) regimes, show finite-time zero events in the total ES. Thus, the exact sector-resolved ES rule is special to \(\mu=0\), but the total ES zero-event diagnostic remains visible under weak sector mixing.

\bibliography{References}

\end{document}